\documentclass[twocolumn]{aastex631}

\usepackage{longtable}
\usepackage{rotating}
\usepackage{afterpage}
\usepackage{booktabs,tabularx}
\usepackage{hyperref}

\usepackage{blindtext}
\usepackage{graphicx}
\usepackage{amssymb}
\usepackage{amsmath}
\usepackage{rotating}
\usepackage{pifont}

\makeatletter
\DeclareRobustCommand{\OmHI}{%
 $\Omega_{\mbox{\scriptsize H\,\check@mathfonts\fontsize\sf@size\z@\selectfont I}}$}
\makeatother

\makeatletter
\DeclareRobustCommand{\HI}{%
 \mbox{H\,\check@mathfonts\fontsize\sf@size\z@\selectfont I}}
\makeatother

\makeatletter
\DeclareRobustCommand{\HII}{%
 \mbox{H\,\check@mathfonts\fontsize\sf@size\z@\selectfont II}}
\makeatother

\makeatletter
\DeclareRobustCommand{\NHI}{%
  $N_{\mbox{\scriptsize H\,\check@mathfonts\fontsize\sf@size\z@\selectfont I}}$}
\makeatother

\accepted{16 April 2026}

\shorttitle{Metallicity Evolution of Damped Ly$\alpha$ Systems Out to $z \sim 5$}

\begin{document}

\title{The Qz5 Survey (II): Metallicity Evolution of Damped Ly$\alpha$ Systems Out to $z \sim 5$}

\author[0000-0002-6505-9981]{M.E. Wisz}
\affiliation{Maria Mitchell Observatory, 4 Vestal St. Nantucket, MA 02554, USA}
\affiliation{Department of Physics, University of California, 5200 Lake Rd, Merced, CA, 95343, USA}

\author[0000-0002-9946-4731]{Marc Rafelski}
\affiliation{Space Telescope Science Institute, 3700 San Martin Drive, Baltimore, MD 21218, USA}
\affiliation{Department of Physics and Astronomy, Johns Hopkins University, Baltimore, MD 21218, USA}

\author[0000-0003-0028-4130]{Grecco A. Oyarz\'un}
\affiliation{Department of Physics and Astronomy, Johns Hopkins University, Baltimore, MD 21218, USA}

\author[0000-0003-2973-0472]{Regina A. Jorgenson}
\affiliation{Department of Physics \& Astronomy, Cal Poly Humboldt, 1 Harpst St., Arcata, CA 95521}
\affiliation{Maria Mitchell Observatory, 4 Vestal St. Nantucket, MA 02554, USA}

\author[0000-0001-6676-3842]{Michele Fumagalli}
\affiliation{Dipartimento di Fisica G. Occhialini, Universit\`a degli Studi di Milano-Bicocca, Piazza della Scienza 3, I-20126 Milano, Italy}
\affiliation{INAF - Osservatorio Astronomico di Trieste, via G. B. Tiepolo 11, I-34143 Trieste, Italy}

\author[0000-0002-9838-8191]{M. Neeleman}
\affiliation{National Radio Astronomy Observatory, 520 Edgemont Road, Charlottesville, VA, 22903, USA}

\author[0000-0002-7738-6875]{J. Xavier Prochaska}
\affiliation{Department of Astronomy \& Astrophysics, UCO/Lick Observatory, University of California, 1156 High Street, Santa Cruz, CA 95064, USA}
\affiliation{Kavli Institute for the Physics and Mathematics of the Universe (Kavli IPMU), 5-1-5 Kashiwanoha, Kashiwa, 277-8583, Japan}
\affiliation{Division of Science, National Astronomical Observatory of Japan, 2-21-1 Osawa, Mitaka, Tokyo 181-8588, Japan}

\author[0000-0001-8415-7547]{Lise Christensen}
\affiliation{Cosmic Dawn Center, Niels Bohr Institute, University of Copenhagen, Jagtvej 128, 2200-N Copenhagen, Denmark}

\author[0000-0002-8081-5343]{Eldon Fobbs}
\affiliation{Texas Tech University, Lubbock, TX 79409}
\author[0000-0003-2344-263X]{George D. Becker} 
\affiliation{Department of Physics \& Astronomy, University of California, Riverside, CA, 92521, USA} 
\author[0000-0002-7054-4332]{Joseph Hennawi} 
\author[0000-0003-0960-3580]{G. Worseck}
\affiliation{Institut für Physik und Astronomie, Universität Potsdam, Karl-Liebknecht-Str. 24/25, D-14476 Potsdam, Germany} 
\author[0000-0003-0389-0902]{Sebasti\'an L\'opez} 
\affiliation{Departamento de Astronom\'ia, Universidad de Chile, Casilla 36-D, Santiago, Chile}

\begin{abstract}

Damped Ly$\alpha$ absorbers (DLAs) are the highest \HI\ column density (\NHI) absorption line systems detected in the spectra of background quasars. DLAs dominate the neutral gas content of the Universe ($\Omega_{\rm HI}$) and are used to measure the metallicity evolution of \HI\ gas. In this work, we introduce a sample of five recently detected DLAs at $z > 4.7$, found in mid to high-resolution spectroscopy from VLT/X-shooter and Keck/HIRES. These DLAs were not pre-selected based on metallicity, enabling an unbiased study of the metallicity of HI gas at $z \sim 5$. We also search for DLAs unbiased in metallicity at $0<z<5.5$ from the literature, we apply a combined correction for dust depletion and $\alpha$-enhancement (assuming no depletion of S, Si, and Zn) to Fe abundances, and we measure the cosmic metallicity evolution of $\alpha$-elements using a linear fit with a slope of$-0.22 \pm 0.05$ dex per unit redshift. For the highest redshift bin, we find an \NHI weighted average of $\langle Z \rangle = -2.00$. This value is $4.4\sigma$ deviant from the trend recovered at $z < 4.7$ and a K-S test comparison gives a $2.4 \sigma$ difference. We conclude that the metallicity of HI gas sharply decreases at $z \sim 5$, in agreement with previous tentative evidence. This sharp decrease may be connected with the onset of the enrichment of galaxies' circumgalactic media or with the end of cosmic reionization, though we cannot exclude that it is driven by small sample statistics.
 
\end{abstract}


\section{Introduction} 
\label{sec: intro}
Our current model of the Universe associates the origin of metals with nucleosynthesis in the stellar components of galaxies (e.g. \citealt{Burbidge1957,croton2006b,vogelsberger2014,schaye2015, peroux2020}). The produced metals are ejected onto the interstellar medium (ISM) in the later stages of stellar evolution, before being later recycled in subsequent star-formation episodes (\citealt{nomoto2013,pillepich2018}). Our current understanding is that not all of these metals are immediately recycled, with some of the metal-enriched gas escaping the ISM and reaching the circumgalactic medium (CGM) in a process known as the baryon cycle: gas and metals cycle in and out of galaxies via inflows and outflows \citep[e.g.][]{TPW2017}. The importance of this cycle in the context of our models of galaxy formation and evolution outlines the need for characterizing the chemistry of gas in both the ISM and CGM of galaxies.

A major challenge encountered by studies of the baryon cycle lies in the characterization of neutral (\HI) gas. With the exception of Ly$\alpha$ fluorescence, \HI\ gas cannot be detected in emission at $z>2$ due to the low luminosity of the 21~cm emission line \citep{fernandez2016,chowdhury2020,chowdhury2021,chowdhury2022}. Probing \HI\ gas at these redshifts requires absorption spectroscopy, often by utilizing luminous quasars (QSOs) as background sources (\citealt{wolfe1986,steidel-sargent1992, wolfe2005}). Luminous emission from the QSO acts as a pinprick of light through foreground clouds of gas and galaxies, which in turn produce Ly$\alpha$ absorption features in the QSO spectrum blueward of the Ly$\alpha$ emission peak (e.g. \citealt{rauch1994}).

Larger troughs of Ly$\alpha$ absorption found lurking in the forest are the signpost of higher column density gas, likely associated with the ISM or CGM of intervening galaxies. The most optically thick class of these large Ly$\alpha$ troughs are the damped Ly$\alpha$ systems (DLAs), which are dense clouds of \HI\ gas with a minimum neutral hydrogen column density of \NHI~$> 2 \times 10 ^{20}$~cm$^{-2}$ (\citealt{prochaska-wolfe1997,wolfe2005}). Because of their high column density, DLAs are known to dominate the neutral gas content of the Universe out to $z \sim 5$ \citep{Prochaska2005}. The \NHI\ threshold for DLAs is such that these systems are shielded from ionization, and thus the computation of their chemical abundances does not require ionization corrections. To measure the chemical abundances and/or metallicities of DLAs, studies often turn to associated spectral absorption lines imprinted in the QSO continuum redward of the Ly$\alpha$ emission peak \citep[e.g.][]{Prochaska2003,Rafelski2014,neeleman2013}. 

Moreover, the availability of DLAs across the whole $0<z<5.5$ redshift range makes them a useful tool for building a timeline of the metal content of the Universe over cosmic time. The cosmic metallicity, $\rm \langle Z \rangle \equiv log_{10}(\Omega_{metals}/\Omega_{gas})-log_{10}(M/H)_{\odot}$ \citep{lanzetta1995}, is the column density weighted mean metal abundance, typically measured in redshift bins of equal number of targets. The evolution of DLA metallicity with redshift has been quantified previously \citep[e.g.][]{Prochaska2003}, often recovering a linear decrease with a slope $\langle Z \rangle = (-0.20 \pm 0.03)\,z - (0.68 \pm 0.09)$, as well as a metallicity `floor' at $\rm [M/H] = -3$ \citep[hereafter R12]{Rafelski2012}. Interestingly, a tentative (yet steep) drop-off from this linear trend with redshift was observed at $z>4.7$ \citep[hereafter, R14]{Rafelski2014}. R14 speculate that this drop-off in metallicity could be associated with changes in the metal enrichment of the gaseous components of galaxies in the post-reionization era. Because of this, there has been an increased focus on observing quasars at high redshifts with the aim of further constraining the properties of DLAs at $z \sim 5$ (\citealt{crighton2015, Poudel2018,oyarzun2025}). 

In this work, we present an update on the cosmic metallicity evolution of alpha elements previously presented in the R12 and R14 studies. This is achieved through the Qz5 survey which provides high resolution spectra of 63 QSOs at $z \sim 5$  intended to double the number of observed DLAs in the literature at $z > 4.7$. However, this survey only yielded five new non-proximate DLAs at this redshift, indicating that the DLA incidence rate at $z\sim 5$ is lower than at $z<5$ \citep{oyarzun2025}. Our main goal is to increase the sample of DLAs at $z > 4.7$, allowing for a more precise measurement of the metallicity of DLAs at this redshift.  In this paper, we present a detailed chemical analysis of this sample of five recently detected DLAs $z > 4.7$ from the Qz5 survey, increasing the total number of known DLAs in this redshift regime from 14 to 18. 

We additionally conduct a comprehensive literature search for DLAs unbiased in metallicity discovered in the last decade across $0<z<5$ since R14. We exclude any DLAs that were pre-selected based on their metallicity, any proximate DLAs, and any sub-DLAs. Finally, we also introduce a novel method to correct metallicity measurements based on metal ions affected by dust depletion (\citealt{jenkins2009, DeCia2016,decia2018, vladilo2018,Konstantopoulou2022,RomanDuval2022ApJ...928...90R,RomanDuval2022ApJ...935..105R,Hamanowicz2024}), when multiple high resolution transitions are not available. The updated $z>4.7$ sample, the comprehensive literature search for DLAs at $0<z<5$, and our approach to correct for $\alpha$-enhancement and dust depletion allows us to present the most robust, up-to-date study of the redshift dependence of the cosmic metallicity of \HI\ gas in the Universe.

The outline of this paper is as follows. In Section \ref{sec: data}, we introduce the Qz5 Survey including the observations of five DLAs at $z > 4.7$. In Section \ref{sec: methods}, we introduce metallicity measurements for the new $z > 4.7$ DLAs and an updated sample of DLAs resulting from a comprehensive literature search over the last decade. We also present our dust depletion and $\alpha$-enhancement correction method. Section \ref{sec: metal evolution} details the metallicity evolution of alpha elements. We present our results confirming a turnover in metallicity at $z \sim 4.7$. Finally, in Section \ref{sec:discussion} we discuss differences in methods of calculating the significance of our results, our approach to dust depletion and $\alpha$-enhancement corrections, and expectations from theory and simulation in the context of our results. We provide a final summary of our work in Section \ref{sec: summary}. 

\section{Data}
\label{sec: data}
In this section we present the survey of high-redshift quasars in which our five $z > 4.7$ DLAs were discovered. We discuss the motivation for conducting the survey, describe the observations with VLT/X-shooter, and report the discovery of the five DLAs, which includes the measurements of their $N_{\rm HI}$ values. 

\subsection{Sample Observations and Data Reduction}

The Qz5 Survey \citep{qz5}, first introduced in our companion paper \citet{oyarzun2025}, is a spectroscopic survey of 63 $z \geq 5$ quasars selected from the SDSS BOSS survey and observed with VLT/X-shooter and Keck/ESI between 2013 and 2018 (PI: Rafelski; Program IDs: 098.A-0111 and 0100.A-0243). The quasars were originally selected purely on redshift and magnitude with the intention of more than doubling the sample of DLAs at $z \geq 4.7$. However, only five intervening DLAs were discovered along the line of sight of the targets. Although we found the incidence rate of DLAs at high-redshift to be much lower than anticipated, we use these new high-redshift targets to measure the neutral hydrogen \HI\ gas density of the Universe ($\rho_{\rm HI}$). We found that $\rho_{\rm HI}$ doubles from $z \sim 5$ to $z \sim 3$. We refer the reader to \citet{oyarzun2025} for details about the observations, data reduction, and in-depth DLA identification methods. 

In this work, we present an analysis of the five DLAs found in \citet{oyarzun2025}. In addition to the VLT/X-Shooter data from \citet{oyarzun2025}, Keck/HIRES data were obtained in October and November 2016, February and October 2017, and June 2018 split between independent programs led by G. Becker (Program Ids: U164Hr and U052) and J.Hennawi (Program Ids: U074 and U189). The spectra we present in this paper for our analysis use a combination of the X-Shooter and HIRES data. Given the higher spectral resolution, we prioritize HIRES data where available, then fill in higher wavelength line transitions using X-Shooter data.

\subsection{Data Processing}
For fitting of the quasar continuum, we used the {\tt lt\_continuumfit} script from the {\tt linetools} package \citep{linetools}. This script allows for manual spline fitting of the quasar continuum around DLA absorption components both blueward (for the \NHI\ fits) and redward (for the metal lines) of the quasar Ly$\alpha$ emission peak. The process of fitting the \NHI\ is presented in detail in \citet{oyarzun2025} and the results are shown in Fig. \ref{fig: nhi}.

\subsection{Identification of DLAs}
\begin{figure*}
    \includegraphics[width = 7.1in]{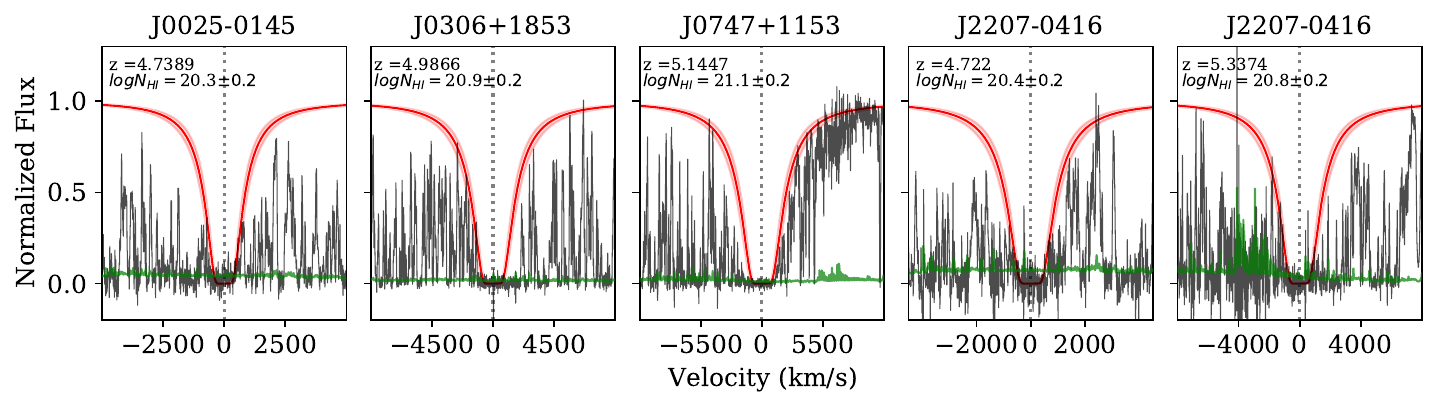}
    \caption{Voigt profile fits for the $\rm N_{HI}$ of each DLA system, ordered by redshift from \citet{oyarzun2025}. The spectrum is plotted in black, with the uncertainty in green. For the Voigt profile fits, the dark red line shows the best fit and the shaded region indicates the uncertainty on the fit. The vertical dashed line marks the velocity centroid, determined by low-ion metal transitions. DLA redshifts and \NHI values are printed at the top of each plot.}
    \label{fig: nhi}
\end{figure*}

The search for DLAs in this dataset is described in detail in a companion paper \citep{oyarzun2025}. In summary, we searched for DLAs in the 63 sightlines across the redshift range $4.5<z<5.6$. Because of the high spectral resolution (R $> 7000$) and signal-to-noise (S/N $\geq 17.0$), the search could be easily conducted visually, which yielded $>40$ DLA candidates. Voigt profiles were fit to the absorbers to estimate their $\rm N_{HI}$. Analysis revealed that only eight candidates satisfied the DLA \HI\ column density threshold, but three were proximate DLAs. Our proximate threshold for this work is $ < 5000 \rm km/s$, and thus we exclude any DLAs that have a velocity separation from the host smaller than this limit. 31 candidates were sub-DLAs. Of the five remaining non-proximate DLAs, one was previously presented in R14, but is also included here for the completeness of the Qz5 Survey. The Voigt profile fits of all five new high-redshift DLAs are shown in Fig. \ref{fig: nhi} and are described in more detail in \citet{oyarzun2025}. 

We note that the DLAs in J0025-0145 at $z = 4.7388$ and J2207-0416 at $z = 5.3374$ appear to have small flux peaks at the bottoms of the DLA troughs. For both of these instances, the apparent flux peaks are well understood as residuals with the skyline subtraction from our reduction methods. For the case in J0025-0145, the peaks are so narrow that they are most likely due to skyline subtraction error and the level of confidence in our telluric corrections. For this system in particular, it is not possible to fit a system that is physical given the column densities we report later on in this work, should the peaks be considered as true flux. The metal lines we measure are consistent with the characteristics and appearance of a DLA. For the case of J2207-0416, it is clear from the error array shown in green in Fig. \ref{fig: nhi}, that the peaks at the redward side of the DLA trough are from sky subtraction. 

We report updated redshifts for the four DLAs in this work measured using the HIRES observations. These values along with other properties are presented in Table \ref{tab:sample}.

\subsubsection{Wavelength Calibration Correction}
VLT/X-shooter has a documented issue with the wavelength calibration that can be present in the VIS and NIR arms, where the wavelength solution outputs wavelengths offset from where they should occur.\footnote{Documentation can be found at:\\ \url{https://www.eso.org/sci/facilities/paranal/instruments/xshooter/doc/XS_wlc_shift_150615.pdf}} This issue is visible in our post-reduction data. To correct for this offset, we use a sky model spectrum from the Cerro Paranal Advanced Sky Model\footnote{\url{https://www.eso.org/observing/etc/bin/gen/form?INS.MODE=swspectr+INS.NAME=SKYCALC}} that was adapted to match the coverage of the VLT/X-shooter data (300nm - 2480nm). We also selected a Gaussian convolving line spread function set to a FWHM of one bin to best match the resolution of VLT/X-shooter. Otherwise, we use the default sky model parameters.

To correct the observed spectra, we performed a one-dimensional cross-correlation analysis between the produced sky model and the error spectrum of our observed data. To determine the offset with sub-pixel precision, we computed the average of the wavelength offsets weighted by the magnitude of the cross-correlation. The offset is then converted to wavelength space for implementation. The offsets are not systematic across each observing run, but they are in agreement with offsets reported by \citet[i.e.; $<1$~\AA\ scales]{Kruhler2015}. Offsets are computed and applied to the relevant wavelength ranges of each X-shooter arm separately. Our data has $0.35$~\AA\ and $0.6$~\AA\ bin sizes for the VIS and NIR arms, respectively. Thus, we do not implement any offsets smaller than those respective values, as the error on the offset is on the scale of the bin size. We do not correct the UVB arm, as lines of interest do not fall in that wavelength coverage for our targets.

\section{Methods}
\label{sec: methods}
In this section we discuss our novel dust depletion and $\alpha$-enhancement correction routine and the processing of spectral data from the VLT/X-shooter and Keck/HIRES follow- up observations of our high-redshift DLA sample. We detail the method for calculating the metallicities of our high-redshift sample and the form of $\alpha \rm -enhancement$ we apply to correct for the use of non $\alpha$ ions ($\alpha$ elements are created via the $\alpha$ process which involves nucleosynthesis via extreme heat, typically generated through the process of stellar nucleosynthesis or synthesized in core-collapse supernovae). Finally, we present a comprehensive sample of DLAs unbiased in metallicity from the literature, with metallicities measured and published in the last decade since R14. 

 \subsection{Dust Depletion and $\alpha$-Enhancement Correction Method}
\label{sec: dust corrections}
A concern in the reporting of observational metallicity measurements is the effect of metal depletion onto dust grains in the ISM. Corrections must be considered for Fe-based metallicity measurements, as the depletion can be significant for this element at high metallicities. Recent work on $\alpha \rm -enhancements$ in the Large Magellanic Cloud (LMC) and Small Magellanic Cloud \citep[SMC,][]{DeCia2024} find that the average $\alpha \rm -enhancement$ is +0.26~dex in the LMC, which is consistent with the measurement reported in R12 (+0.27~dex) for low-metallicity DLAs. There are multiple approaches for attempting to account for this depletion, including the use of Galactic absorption lines \citep{DeCia2016} and/or observations from the LMC and SMC \citep{roman-duval2019, roman-duval2021, roman-duval2022, jenkins2017, tchernyshyov2015, decia2018, Konstantopoulou2022, konstantopoulou2024, Hamanowicz2024} to build a depletion model for different elements. These approaches seem particularly well suited for low redshift DLAs and can be stretched up to $z \sim 2-3$. However, applying depletion corrections at $z>4$ for such a sample that we build in this work is unprecedented because it would require measurements of multiple line transitions that are often redshifted in to the NIR and have small oscillator strengths (i.e. Zn) which makes it extremely difficult to measure when the metallicities are low. Here, we use S as one of the primary elements on which we compute the depletion correction as it is only mildly depleted and is a volatile element. Thus, we are able to compare these results with other $\alpha$ elements at low metallicities especially since we expect that the depletion of low metallicity DLAs is very small.  

Additionally, \citet{velichko2024} investigate $\alpha \rm -enhancements$ in DLAs for high and low $\Delta v_{90}$ samples. They find an average value within $1 \sigma$ of the R12 measurement for the high sample, and within $2\sigma$ for the low sample, however we cannot necessarily compare our results to this work as we are unable to measure $\Delta v_{90}$ for all of the sources in this work. In our combined correction for both dust depletion and $\alpha$-enhancement applies an $\alpha$-enhancement as a minimum correction, which is consistent with the work of \citet{velichko2024}. 

For comparison, we also check our S based corrections with a similar correction based on Zn in Appendix \ref{sec: appendix_a}. Zn is a non-refractory element and would in principal complement the S based corrections. However, we note that the use of Zn is complicated by its production method which is dependent on the host galaxy star-formation history as it likely originates in massive stars \citep{Woosley1995, Hoffman1996, Umeda2002, Fenner2004}. In particular, the dominant nucleosynthetic site of Zn production at low metallicities is believed to be different than at high metallicities, resulting in Zn only being a good tracer of Fe at high metallicities \citep{Berg2015b, Umeda2002, Nissen2007}. This metallicity dependence likely results in Zn appearing to act like an $\alpha$ element in some DLA samples (R12) and makes it challenging to use for depletion corrections over a range of metallicities. We therefore opt to utilize S in our empirical correction while leaving Zn (which has better number statistics) as a general check on our S corrections, acknowledging its limitations. We complete an analysis of our results using Zn as the basis for our correction instead of S, and we note that the primary results of this work hold true.

\begin{figure*}
    \centering
    \includegraphics[width = 7.1in]{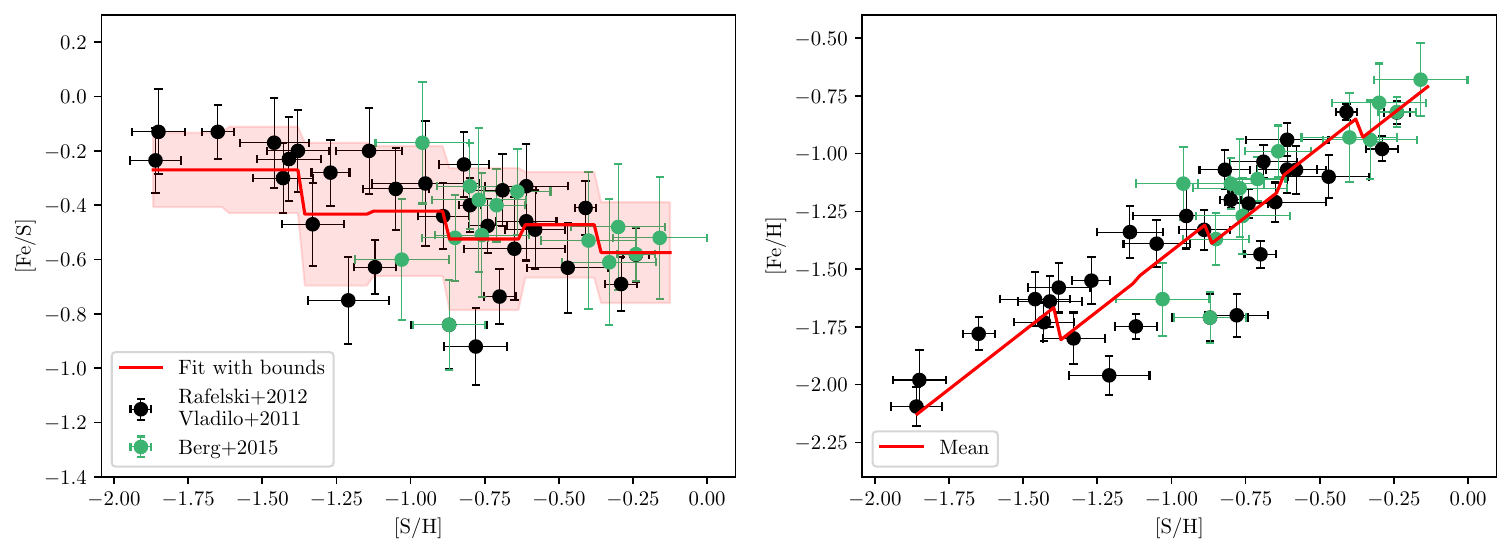}
    \caption{{\emph Left:} Measured gas phase abundances of [S/H] and [Fe/S] according to the studies by R12, \citet{vladilo2011}, and \citet{Berg2015a}.  The vertical and horizontal error bars are uncertainties on the respective measures. After binning the data, the mean relationship between [S/H] and [Fe/S] is shown in red, with uncertainties shown as red shading. These uncertainties are propagated throughout for the measurements we correct using this relationship. This mean relationship is used as a non-parametric dust depletion and $\alpha$-enhancement correction for all Fe-based metallicities (Section \ref{sec: dust corrections}). {\emph Right:} Parameterized relationship from the left panel fit to [S/H] vs [Fe/H] parameter space.}
    \label{fig:s_fe}
\end{figure*}

To implement a dust depletion and $\alpha$-enhancement correction to the metallicities of our DLAs that are measured using Fe lines, we opt for a novel, data-driven method. We turn to the observed relation between elements that are less affected by dust depletion (in this case [S/H]) and the depletion affected measurements of [Fe/H]. We binned the measurements of [Fe/S] against [S/H] in R12 to obtain a non-parametric relationship between these two quantities (left panel of Figure \ref{fig:s_fe}). During the fitting process, this relationship, which we denote $\delta_{S} = \delta_{S}(\rm [S/H])$, was set to a maximum value of  
\begin{equation}
\delta_{S}(\rm [S/H]) <-0.27,
\end{equation}
This constraint is motivated by the $\alpha$-enhancement correction. This threshold only impacts the lowest [S/H] bin, and the resulting fit can be observed in Figure \ref{fig:s_fe}.

We can utilize $\delta_{S}$ to estimate an Fe-based metallicity for every DLA. We refer to this average, corrected metallicity $\rm[Fe/H]_{corr}$ and to the uncorrected original measurement $\rm[Fe/H]$. These two quantities are related by
\begin{equation}
\rm[Fe/H]_{corr} = [Fe/H] - [Fe/S]
\end{equation}
Ideally, we would utilize measurements of the dust depletion [Fe/S] for every DLA to obtain the true $\rm[Fe/H]_{corr}$. Because this is not possible for our dataset, we utilize the fitted $\delta_{S}$ to obtain an average dust depletion correction ($\rm [Fe/S]^*$) instead
\begin{equation}
\rm[Fe/H]_{corr} = \rm [Fe/H] - \rm [Fe/S]^*, 
\end{equation}
where $\rm [Fe/S]^*$ is the solution to the equation
\begin{equation}
\label{depletion_eq}
\rm [Fe/S] = \rm [Fe/H] - \rm[S/H] = \rm[Fe/H] - \delta^{-1}_{S}([Fe/S]), 
\end{equation}
where $\delta^{-1}_{S}$ corresponds to the inverse function of $\delta_{S}$, such that it yields a value for [S/H] for an input [Fe/S]. This means that for every measured value of [Fe/H] there is a unique value $\rm [Fe/S]^*$ that solves Equation (\ref{depletion_eq}).

We apply this correction to all DLAs with reported metallicity values that use Fe lines, uncorrected for $\alpha$-enhancements that may have been previously applied in the values obtained from the literature. The uncertainty on the metallicity measurement with the sulfur dust depletion and $\alpha$-enhancement correction is additionally corrected via addition in quadrature of the original uncertainty and the uncertainty on the dust depletion and $\alpha$-enhancement correction trend. For low-metallicity targets where the dust depletion and $\alpha$-enhancement correction  needed becomes quite small and [S/Fe] is dominated by $\alpha$-enhancement, we use the error on the $\alpha$-enhancement measurement reported in R12 of $\pm 0.02$ dex. For all remaining targets, we use the standard deviation of the distribution of the measured [Fe/S] value and the predicted value (value at the red line) for the data in the left panel of Fig. \ref{fig:s_fe}. These corrected uncertainties are also listed in Table \ref{tab:sample}. We additionally compute the depletion based on Zn in Appendix \ref{sec: appendix_a}, and show that using Zn instead of S for the depletion corrections does not change the cosmic metallicity evolution fit. Finally, we plot the trend we derived to fit [S/H] vs. [Fe/S] over the gas-phase abundance space ([S/H] vs. [Fe/H]), and we see that the trend follows this data well. This alternate parameter space acts as a check for our depletion correction routine and yields an individual [Fe/H] value for every [S/H] measurement. 

\subsection{Metallicity Measurements}
We use the apparent optical depth method \citep[AODM]{SandS1991} to measure all atomic column densities for our sample of DLAs. We follow the line preference hierarchy laid out in R12 to determine which lines are best to use for the metallicity measurement for each target, based on which lines are available and the current understanding of depletion levels. O, S, and Si are of top priority to use because they contribute to larger fractions of the heavy element mass density of stars and are formed via similar processes in supernovae. Zn is the next best ion to use if O, S, or Si are not available because it is non-refractory and the two strongest transitions often lie outside the Ly$\alpha$ forest. Finally, Fe and Ni can be used when none of the previous ions are available. For a more detailed explanation, see Section 3.2 in R12. 

Where HIRES follow-up data is available, we prefer measurements from those spectra over the X-shooter data because of the higher spectral resolution. This dramatically improves the quality of our measurements, as some lines that appear as detection at the resolution of X-shooter appear saturated in the higher resolution HIRES spectra, keeping in mind that AODM does not work for saturated lines. Thus, we report lower limits where necessary. The primary observed abundance value used for each metallicity is reported in Table \ref{tab:sample}, as well as the dust depletion and $\alpha$-enhancement corrected value based on sulfur that we use for our analysis. 

Figures \ref{fig:J0025}, \ref{fig:J0306}, \ref{fig:J0747}, \ref{fig:J2207_2}, and \ref{fig:J2207} show metal line transitions for the five new $z > 4.7$ DLAs with metallicity measurements, where the range of the x-axis is set according to the apparent velocity width of the transition. The colored regions denote the applicability of the respective line for the metallicity measurement, with blue indicating the line used for the reported metallicity, green marking a detection, purple indicating an upper limit, and red corresponding to a lower limit/saturated line. Gray regions indicate where the velocity region was truncated due to contamination. The dotted red line at the bottom of each panel indicates the uncertainty and the dashed gray vertical and horizontal lines indicate the velocity centroid of the DLA and the continuum level, respectively.

\subsubsection{J0025-0145 at $z_{abs} = 4.73289$}
\begin{figure*}
    \centering
    \includegraphics[width = 6.9 in]{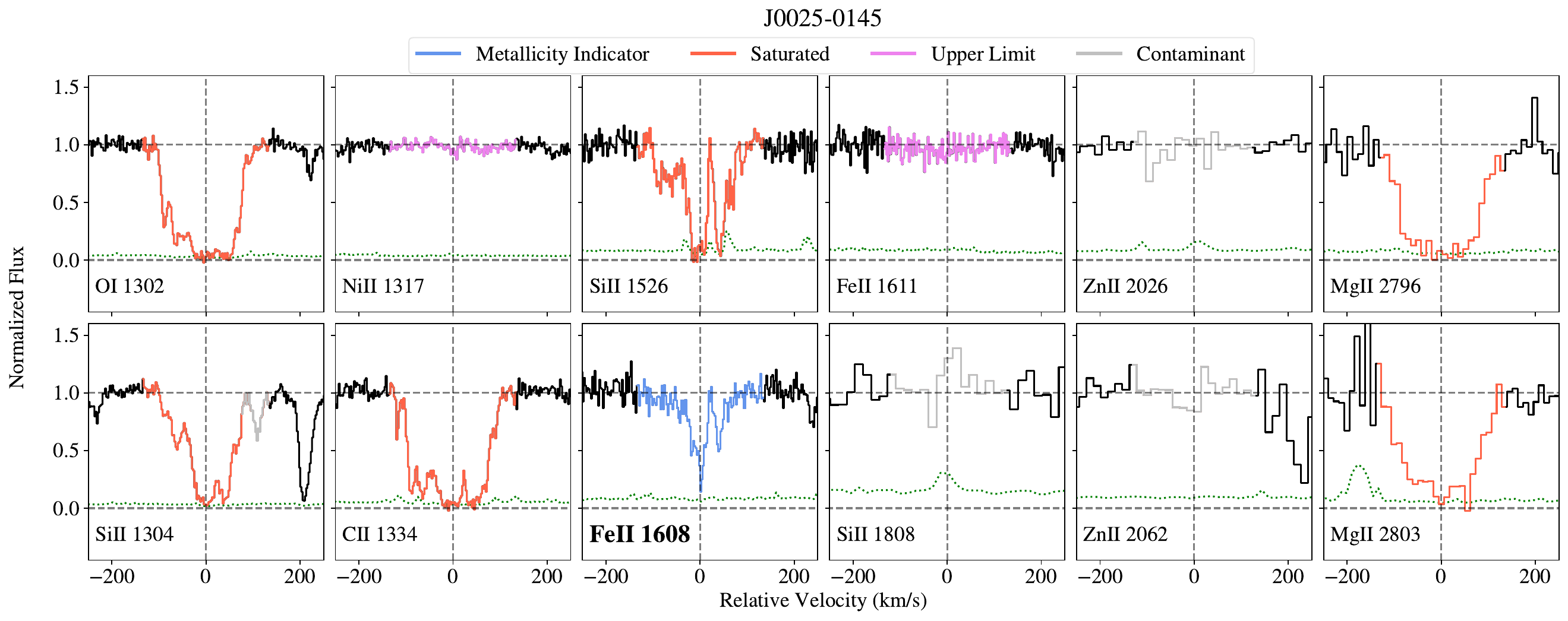}
    \caption{Velocity profiles for the primary metal line transitions for J0025-0145. Metal lines through FeII 1611 are plotted and measured using the HIRES spectrum. The remaining lines plotted are from the X-shooter spectrum. SiII 1808, ZnII 2026, and ZnII 2062 have contamination at the redshift of the DLA (i.e.\ contaminated absorption and/or sky line residuals), and are therefore colored in gray. We use  the FeII 1608 from HIRES to obtain $[{\rm Fe/H}] = -1.14 \pm 0.20$ as the primary metallicity, with a +0.27 dex $\alpha$-enhancement applied, per R12. The measurement from SiII 1304 is higher than this [Fe/H] value, thus we believe that the FeII measurement is more depleted than we measure using our dust depletion and $\alpha$-enhancement correction routine.}
    \label{fig:J0025}
\end{figure*}

From HIRES, we measure ${\rm [Fe/H]} = -1.14 \pm 0.20$ $(\log N_{Fe} = 14.34 \pm 0.02)$ from FeII 1608 and ${\rm [Si/H]} > -0.93 \pm 0.20$ $(\log N_{Si} = 14.88 \pm 0.01)$ from SiII 1304. Velocity profiles for relevant metal line transitions are shown in Fig. \ref{fig:J0025}. Given that the [Fe/H] measurement with the $\alpha$-enhancement correction applied is more metal poor than the lower limit provided by SiII 1304, the FeII measurement is likely depleted. From the other ion lines available in the X-shooter data, we obtain an upper limit of ${\rm[Zn/H]} < -0.2 \pm 0.30$ $(\log N_{Zn} = 12.73 \pm 0.22)$ using ZnII 2026, which is in agreement with our FeII measurement and SiII limit from HIRES, however ZnII 2026 is contaminated in the X-shooter data, so this limit is unreliable. SiII 1808 is blocked by a skyline in the spectrum, so we are unable to measure this line. Additionally, the component at ZnII 2062 is artificial, as we do not see a detection in ZnII 2026, which has a higher oscillator strength. This DLA is previously reported in \citet{huyan2023} as a super-solar absorber based on abundance measurements of OI 1302 and CII 1334 from a MIKE spectrum, however we measure those lines using a much higher resolution HIRES spectrum and observe them to be saturated. We remeasure these lines using our HIRES data and yield $\rm [O/H] > -1.80$ and $\rm [C/H] > -1.64$, respectively. These saturated lines yield lower limits than the ones we report above and are thus in agreement with the final metallicity we report for this target. Thus, those line measurements in \citet{huyan2023} are unreliable due to the level of saturation, even at high resolution. 

\subsubsection{J0306+1853 at $z_{abs} = 4.9866$}
\begin{figure*}
    \centering
    \includegraphics[width = 6.9 in]{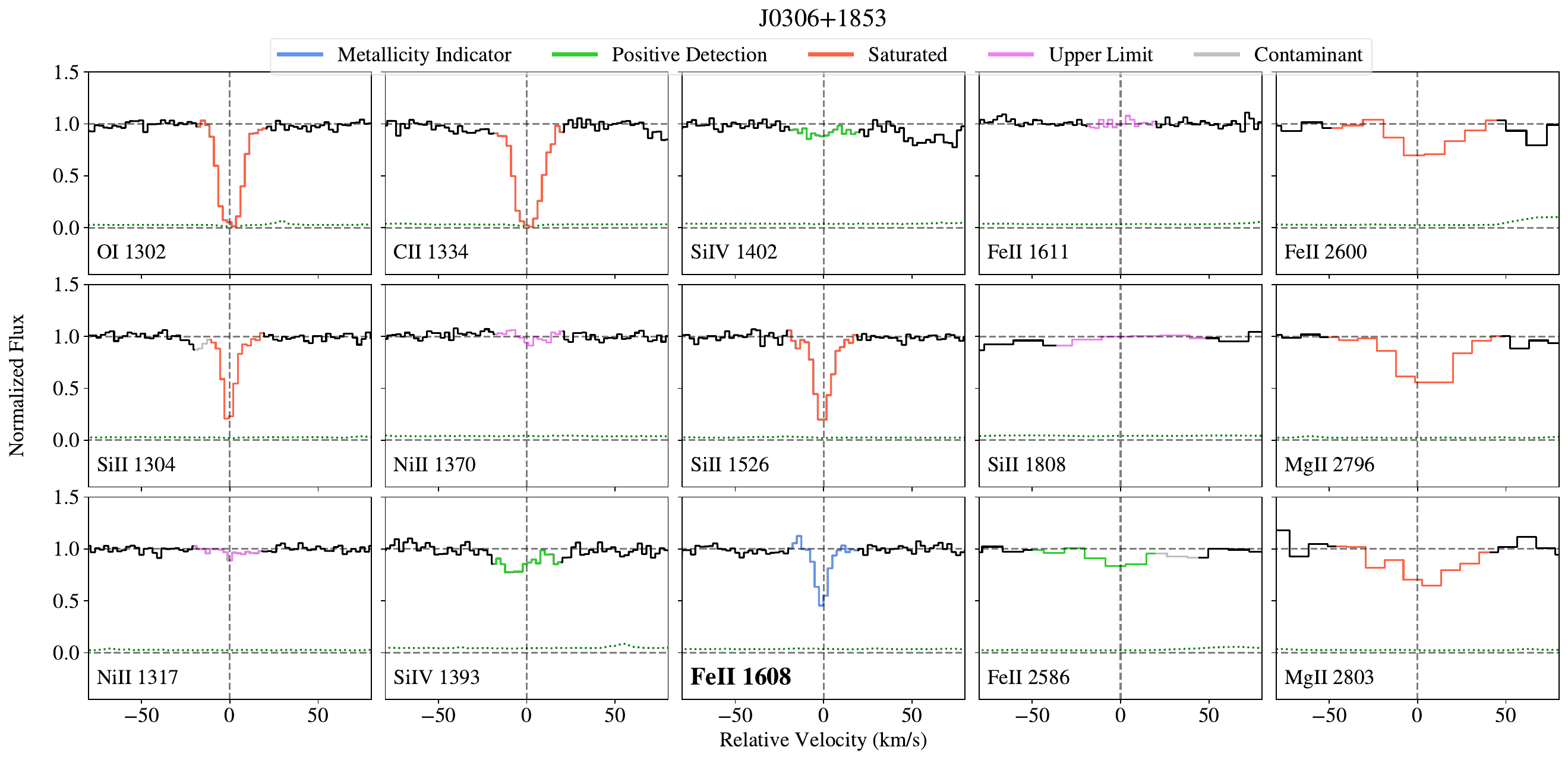}
    \caption{Velocity profiles for the primary metal line transitions for J0306+1853. Metal lines through FeII 1611 are plotted and measured using the HIRES spectrum and the remaining high-ion metal lines are from the X-shooter spectrum. We use FeII 1608 from HIRES to obtain a metallicity of [Fe/H] $= -2.74 \pm 0.20$ as the primary metallicity, with a +0.27 dex $\alpha$-enhancement applied, per R12. Similar to J0025-0145, we measure a slightly higher, but mostly consistent [M/H] value from SiII 1304 in HIRES compared to the [Fe/H] measurement for this DLA, thus we believe that FeII is slightly depleted and that SiII is not very saturated, if at all. }
    \label{fig:J0306}
\end{figure*}

We measure the abundance of iron to be [Fe/H] $= -2.74 \pm 0.20$ $(\log N_{Fe} = 13.33 \pm 0.04)$ from FeII 1608 in the HIRES data. This is higher than a SiII 1304 measurement of [Si/H] $> -2.78$ $(\log N_{Si} = 13.53 \pm 0.01)$, even with the $\alpha$-correction, so we conclude that SiII 1304 is likely mildly saturated at the resolution of HIRES. The metallicity given by SiII 1304 is very consistent with the dust depletion corrected metallicity measurement for this DLA, additionally pointing to the SiII 1304 measurement being only mildly saturated, if at all. Velocity profiles for the relevant metal line transitions are shown in Fig. \ref{fig:J0306}. This is the most metal-poor DLA in our new sample. 

\subsubsection{J0747+1153 at $z_{abs} = 5.1447$}
\begin{figure*}
    \centering
    \includegraphics[width = 6.9 in]{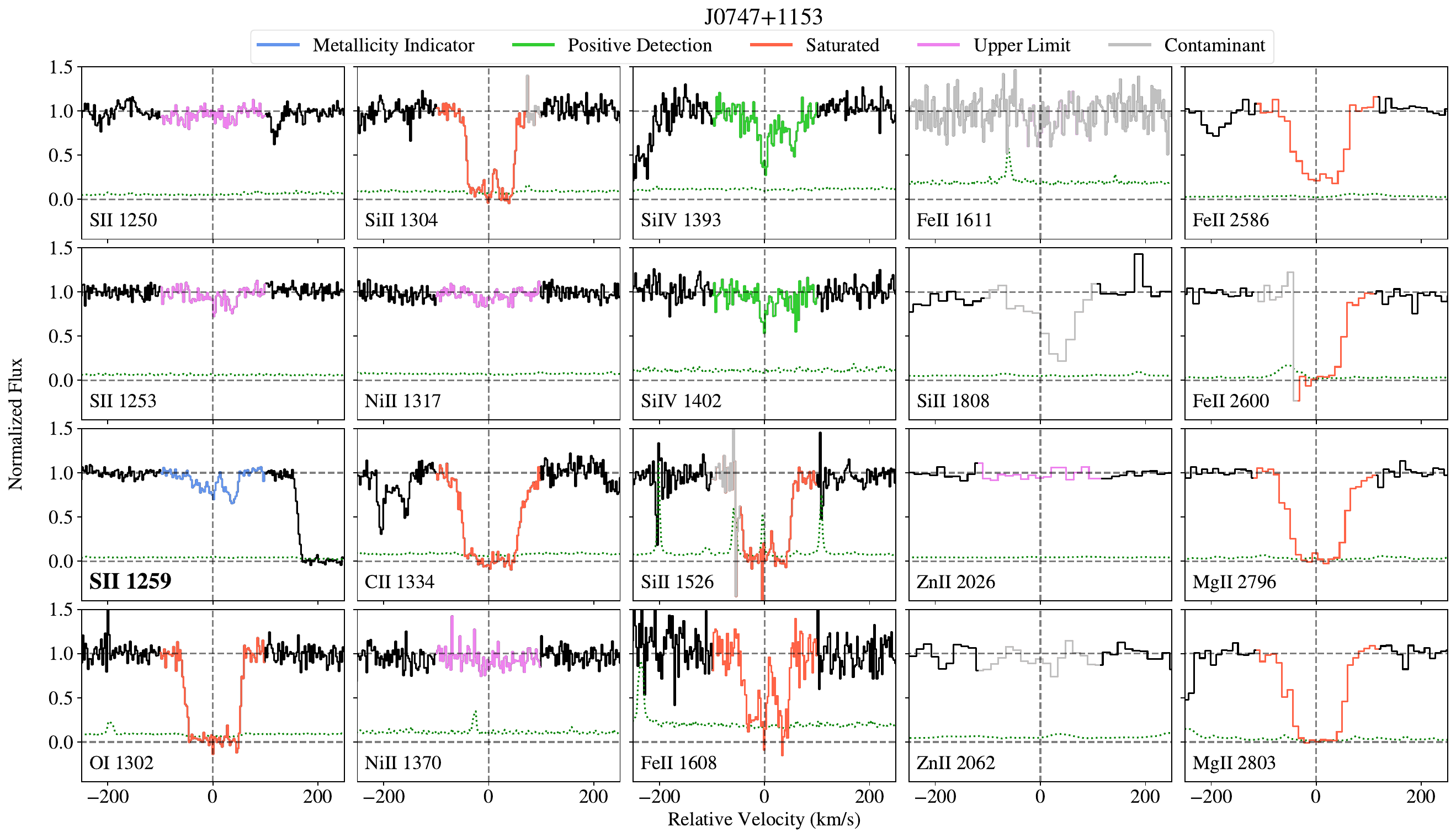}
    \caption{Velocity  profiles for the primary metal line transitions for J0747+1153. Metal lines through FeII 1611 are plotted and measured using the HIRES spectrum and the remaining line transitions are from the X-shooter spectrum. SiII 1808 and ZnII 2062 are contaminated and are thus colored gray. SiII 1808 is actually blended with a MgII absorber at z = 2.973. We use SII 1259 from HIRES to obtain a metallicity of [S/H] $ = -1.71 \pm 0.20$ as the primary metallicity of this DLA.}
    \label{fig:J0747}
\end{figure*}

We measure [S/H] $= -1.71 \pm 0.20$ $(\log N_S = 14.54 \pm 0.02)$ from SII 1259 in the HIRES spectrum. We do not include measurements from the other low ion sulfur lines with lower oscillator strengths due to noise. The SiII 1808 line is blended with a MgII 2796 line from a strong MgII absorber in the line-of-sight at z = 2.9734. Velocity profiles for the relevant metal line transitions are shown in Fig. \ref{fig:J0747}. This DLA and its metallicity were previously reported in \citet{bielby2020} and our measurements are in agreement with what they found. 

\subsubsection{J2207-0416 at $z_{abs} = 4.722$}
\label{sec: J0027 @ z = 4.722}

\begin{figure*}
    \centering
    \includegraphics[width = 6.9 in]{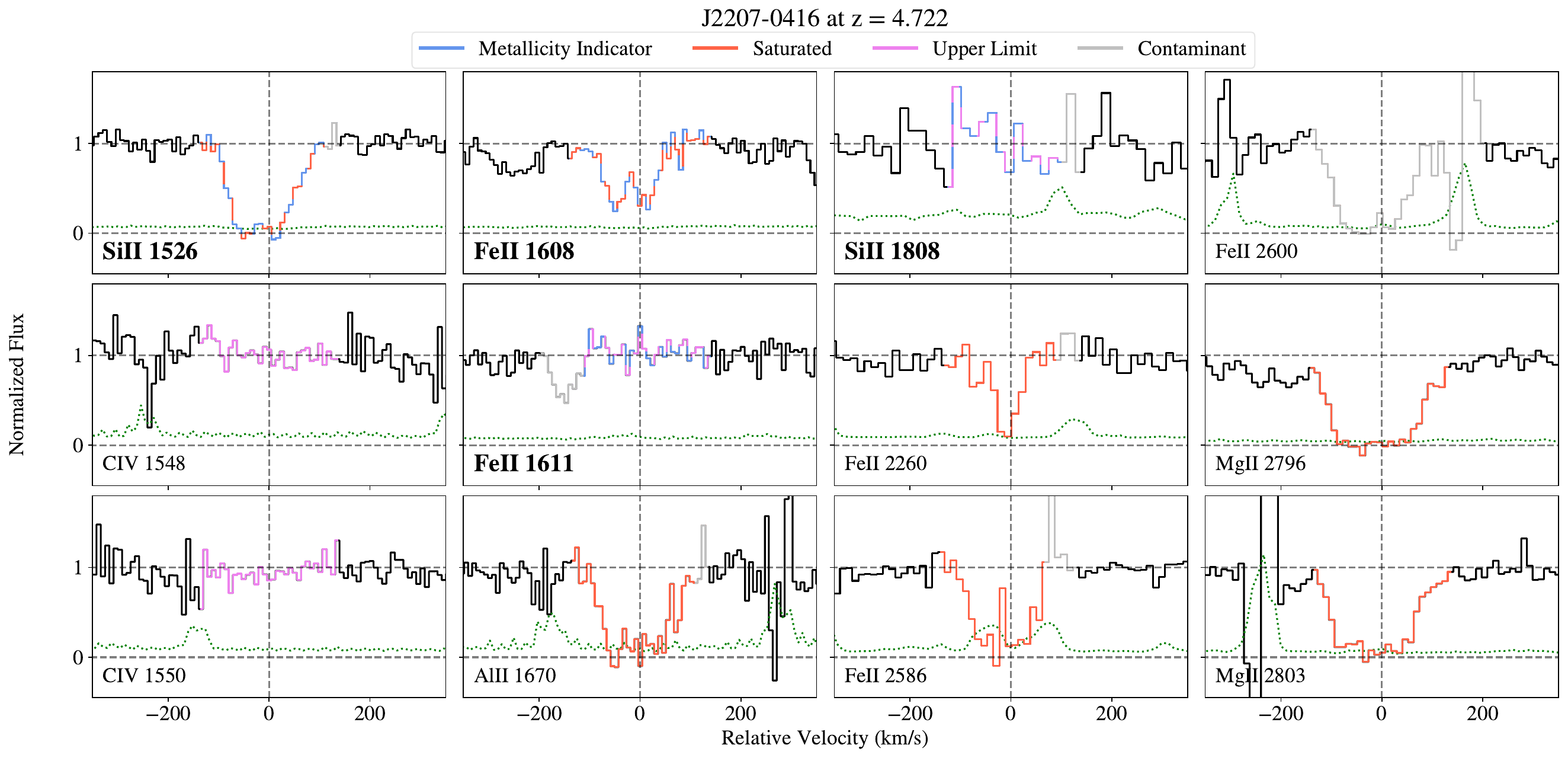}
    \caption{Velocity profiles for the primary metal line transitions for the DLA at $ z= 4.722$ towards J2207-0416. Blue colored lines with alternating colors indicate that we report measurements for the line transitions in this work, while acknowledging that they are limits. We are unable to measure any transitions in the HIRES, thus only the X-shooter spectrum is shown here. We do not report any measurements from the HIRES spectrum. }
    \label{fig:J2207_2}
\end{figure*}

Due to the short wavelength coverage of HIRES paired with the redshift of this DLA relative to the QSO, many lines of interest are in the forest or unrecoverable because they fall in poor sky-line regions. Thus we measure a series of limits from our X-shooter spectrum. We measure lower limits of $\rm [Fe/H] \geq -0.93$ $(\log N_{\rm FeII1608} \geq 14.65)$ and $\rm [Si/H] \geq -1.10$ $(\log N_{\rm SiII1526} \geq 14.81)$ with $3\sigma$ uncertainty. We additionally measure upper limits of $\rm [Fe/H] \leq -0.45$ $(\log N_{\rm FeII1611} \leq 15.13)$ and $\rm [Si/H] \leq -0.27$ $(\log N_{\rm SiII1808} \leq 15.64)$, also with $3\sigma$ uncertainties. We can reasonably estimate that the true metallicity of the DLA lies between $ -1.10 \leq \rm [M/H] \leq -0.27$ based on these limits. We also believe that FeII 2600 is contaminated due to a mismatch between the oscillator strength and the strength of the feature shown in Fig. \ref{fig:J2207_2}, especially when compared to the other identified Fe line transitions. Thus, we do not include it in any Fe based metallicity measurements for this target. Velocity profiles for the relevant metal line transitions are shown in Fig. \ref{fig:J2207_2}.

\subsubsection{J2207-0416 at $z_{abs} = 5.3374$}
\begin{figure*}
    \centering
    \includegraphics[width = 6.9 in]{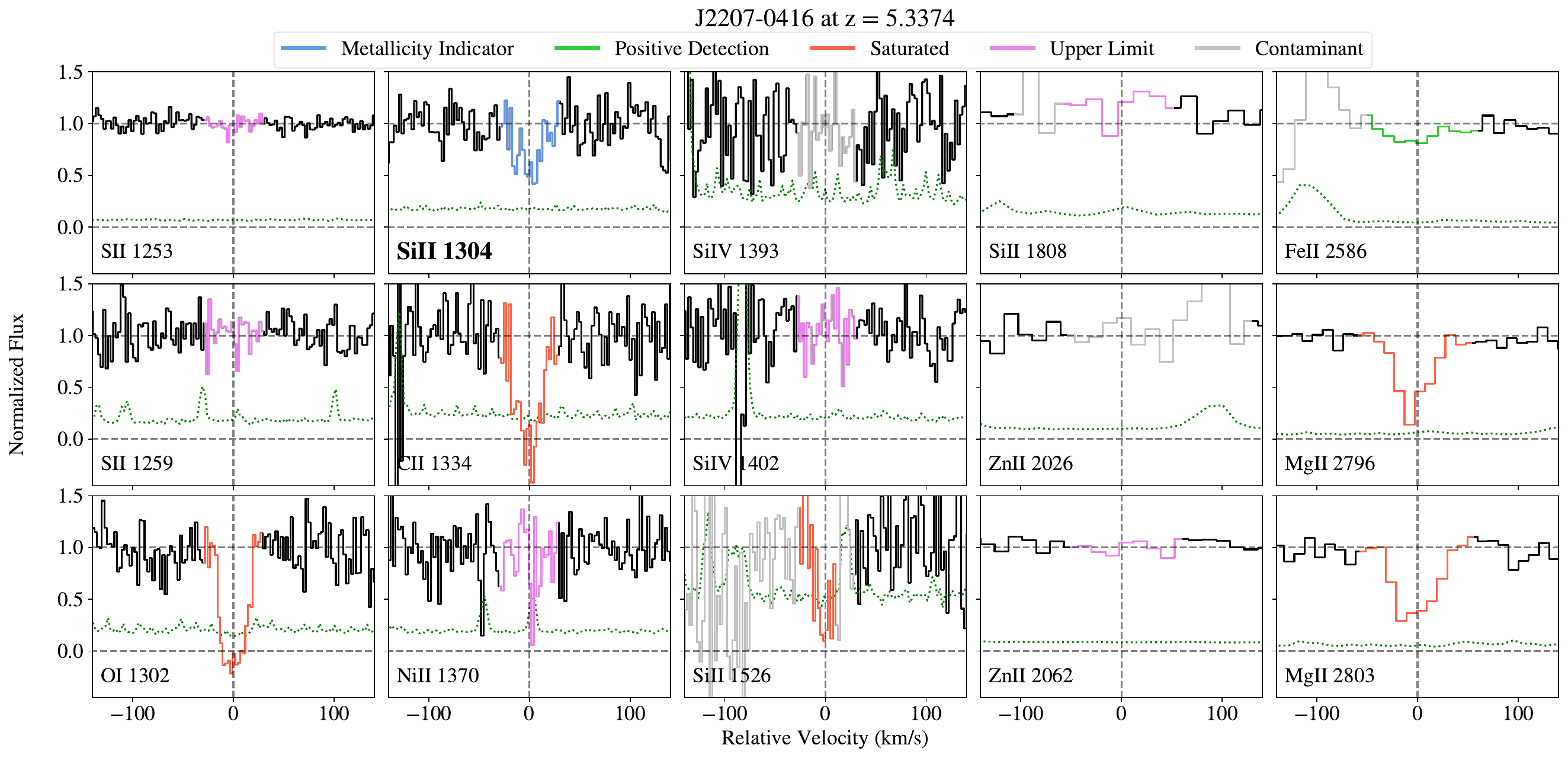}
    \caption{Velocity profiles for the primary metal line transitions for the DLA at $z = 5.3364$ towards J2207-0416. Metal lines through SiII 1526 are plotted and measured using the HIRES spectrum and all remaining lines are from the X-shooter spectrum. }
    \label{fig:J2207}
\end{figure*}

From the HIRES data, we measure [Si/H] $= -2.56 \pm 0.22$ $(\log N_{Si} = 13.75 \pm 0.08)$ from SiII 1304. We include SiII 1526 in this measurement as well, but the overall metallicity value does not change with this inclusion. Thus, SiII 1526 is not very saturated, if at all at the resolution of HIRES. However, from the X-shooter data, the SiII composite metallicity measurement of SiII 1304 and SiII 1526 is lower than the measurement of just SiII 1304, so SiII 1526 is saturated in X-shooter for this target. Additionally we measure a lower limit from OI 1302 of ${\rm [O/H]} > -2.78 \pm 0.98$ $(\log N_O = 14.71 \pm 0.96)$ and an upper limit of ${\rm [S/H]} < -1.91 \pm 0.30$ $(\log N_S = 14.04 \pm 0.22)$ from SII 1259. Finally, from X-shooter, we measure [Fe/H] $=-2.56 \pm 0.22$ $(\log N_{Fe} = 13.42 \pm 0.08)$ from FeII 2586. Velocity profiles for the relevant metal line transitions are shown in Fig. \ref{fig:J2207}. 

\subsection{Literature Search}
\label{sec: lit search}
In addition to our sample of newly observed DLAs at high redshift, we conduct a comprehensive literature search for unbiased DLAs (not pre-selected based on the existence of metal lines). Additionally, for all metallicity measurements, we require a minimum resolution of R $> 7000$ to minimize hidden saturation of line transitions. We start with a base sample consisting of the full samples compiled in R12 and reported in R14, and all 25 DLAs from \citet [hereafter, L13]{Lehner2013}. We include L13 in our base sample, as they reported updated and corrected metallicity measurements for 16 DLAs first reported in R12 (except for Q0256+0110, which is a duplicate from \citet{Nestor2008} of J0256+0110 from \citet{Peroux2006}). For the 16 duplicate DLAs between R12 and L13, we only consider the measurements from L13. We also remove duplicates internal to R12 and remove J1345+2329 from the R14 sample as it was identified as a proximate DLA in \citet{oyarzun2025}. For the full list of references from the sample compiled in R12, please refer to \citet{Rafelski2012}. Our threshold for a DLA to be proximate to the quasar for this work is $ < 5000 \rm km/s$ and thus we exclude any DLAs from the literature that lie below this limit. This threshold is higher than previous works but is consistent with \citet{oyarzun2025}. Due to this change, we do remove 21 DLAs previously included in R12 and R14, one of which is at $z = 4.821$. We make the increase to $5000 \rm km/s$ to conservatively avoid the range in which line transitions could become blended with those of the quasar. 

While we are principally focused on increasing the sample statistics in the high-redshift regime, we include DLAs found at any redshift, published after R12 and prior to the publication of this work. We create sample of DLAs unbiased in metallicity to add to the overall sample of DLAs. DLAs excluded on the basis of bias have been pre-selected for some characteristic or condition such as metallicity or the presence of a type of nearby feature or galaxy. We limit our search to DLAs that are not pre-selected based on strong or weak metallicities. We also exclude any sub-DLAs, LLSs, SLLSs, and proximate DLAs from this work. We consider a total of 46 publications in our literature search (including R12, R14, and L13), and include DLAs from the following seven publications:

\begin{itemize}
 \item We include all DLAs reported in \citet{Fynbo2017}, \citet{Bashir2019}, \citet{Oliveira2014}, and \citet{Kanekar2014} (5 DLAs total).
 \item \citet{Berg2016} -- We include all DLAs listed in their Table 1 except for those identified as proximate. We exclude J0034+1639 at $z = 3.752$, J0747+2739 at $z = 3.424$, and J0955-0130 at $z = 4.024$ because the lines used for the reported metallicity fall either in the UVB or NIR arm of X-shooter, which do not meet our resolution threshold. We finally exclude J1723+2243 because the line used to calculate the reported metallicity is saturated and all other lines are reported as limits. (32 DLAs total). 
 \item \citet{Fumagalli2014} -- We include all DLAs listed in their Table A1, except for the DLAs reported in field G1 and G7 because they are sub-DLAs, and the six DLAs with metallicities bracketed by Si/Zn limits. All other reported DLAs in this paper have been previously included via earlier surveys (15 DLAs total). 
 \item \citet{Srianand2016} -- The absorber at $z = 2.7377$ is excluded because it is a sub-DLA ($N(HI) = 20.07 \pm 0.05$) and the absorber at $z = 3.2477$ is excluded from our sample because it is proximate to the host quasar. All other absorbers in the line of sight of J2358+0419 are included (3 DLAs total).
 \item \citet{Heintz2018} -- The dusty DLA at $z = 2.428$ towards the QSO KV-RQ1500-0031 is included. This QSO also has a strong MgII absorber in the line of sight at $z = 1.603$ (1 DLA total). 
 \item \citet{Kanekar2018} -- The DLAs towards J1553+3548 and J1415+1634 are excluded because they are sub-DLAs (for both, $N(HI) = 19.7$). J1415+1634 is additionally biased due to pre-selection for the presence of a low-mass, gas-rich galaxy within 15 kpc of the QSO sight-line, as stated in the paper. The same pre-selection bias is applied to J0951+3307, and this DLA is therefore also not included. J0930+2845 is excluded as well because its reported metallicity is bracketed by limits, and we exclude limits from our sample to avoid biases. Finally, J1619+3342 is already included in \citet{Lehner2013}. All other DLAs mentioned are included (3 DLAs total). 
  \item \citet{Huyan2025} -- We exclude most of this sample due to many of the targets reported being either sub-DLAs or proximate DLAs. We also exclude their measurement for J0025-0145 as it is already in our sample. For the DLAs that we do include, we take the SiII value they report in their Table 3, as we believe that the oxygen values they report are saturated based on the relative line strengths. We do include J1129-0142, both DLAs towards J1336-1830, and J2328+0217. We also increase the uncertainty on the \NHI measurements to $\pm 0.2$ as we believe that the continuum fits they show are consistently lower than where we would place them and the difference can systematically increase the metallicity measurement (4 DLAs total). 
 \item \citet{DeCia2016} -- It is unclear whether the selection of some of the DLAs in the sample from \citet{DeCia2016} satisfy our criteria. In order to provide as clean a sample as possible, we therefore opt to not include this sample in our paper.
\end{itemize}
For the full list of the publications we considered and reasons for exclusion, see Appendix \ref{sec: appendix_c}. 

Based on the results of our literature search (both publications included and excluded), we find that most of the DLAs that have been discovered in the last decade since R14, do not fall within the high-redshift regime of interest ($z > 4.7$). Many DLAs discovered in the last decade at high redshifts are sub-DLAs ($ 19.0 < {\rm N(HI)} < 20.3$). For example, there is one from \citet{Poudel2018} that is previously included in R14, duplicates and sub-DLAs from \citet{Poudel2020}, and a proximate DLA from \citet{DOdorico2018} that we exclude from our sample. This is consistent with the result from \citet{oyarzun2025}, that the incidence rate of DLAs at this redshift range is low. Although the Qz5 Survey was designed to measure a large sample of DLAs at $ z > 4.7$, only five were discovered due to the lower than expected incidence rate. 

Finally, we consider DLAs found in gamma-ray burst (GRB) host galaxies, as some have previously been found to host DLAs \citep{cucchiara2015}. Most GRBs DLAs are associated with the GRB host galaxy itself and therefore may be biased due to the proximity, so we exclude those from our sample. Of the two single GRB events which have an intervening DLA in the line-of-sight \citep{thone2013, sparre2014}, one is proximate and one has too low resolution due to the slit size of the observations to meet our thresholds for this literature sample. Although GRB DLAs have the aforementioned biases that exclude them from the analysis in this work, it is interesting to note that our global metallicity evolution trend is comparable to that in \citet{heintz2023} observed for GRB DLAs.

\section{Metallicity Evolution}
\label{sec: metal evolution}
\begin{figure*}
    \includegraphics[width = 7.1 in]{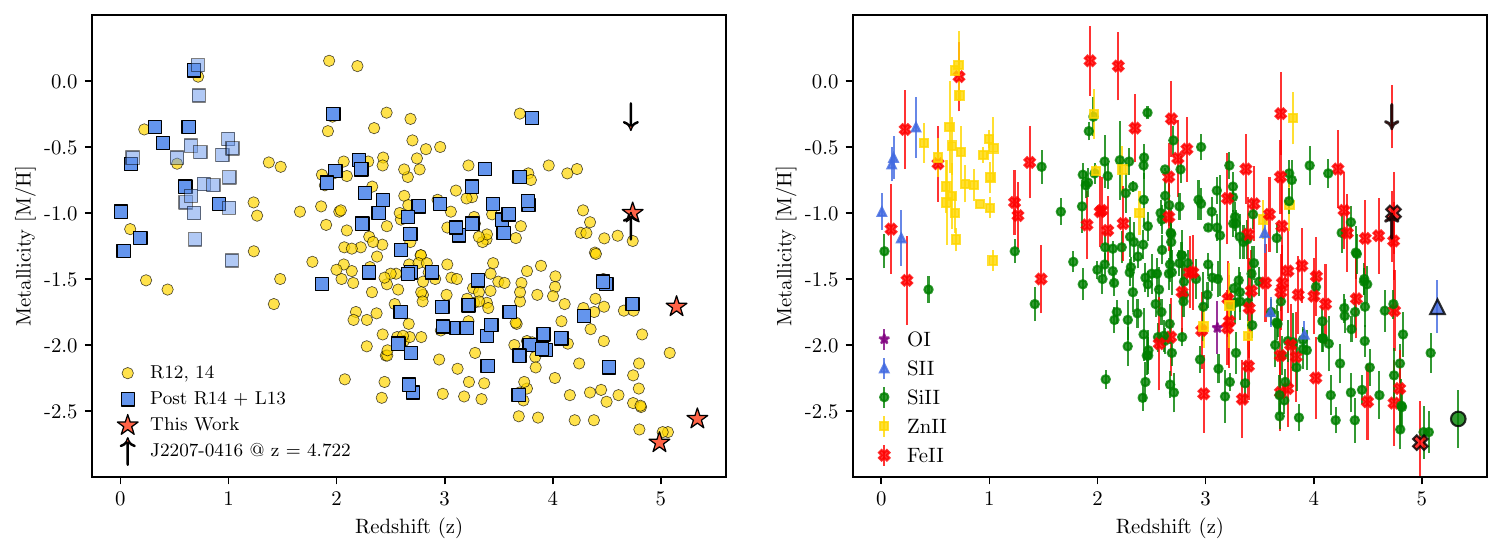}
    \caption{{\emph Left:} Metallicity as a function of redshift of our full updated DLA sample. Previous work from R12 and R14 are shown in yellow circles in the background DLAs from the literature published post R14 are plotted as blue squares, including the L13 DLAs. The DLAs that were previously reported in R12 and updated in L13 are plotted with higher transparency. The four new $z > 4.7$ DLAs from this work are plotted as red stars. {\emph Right:} DLA metal abundance [M/H] vs redshift for all 291 DLAs. The symbols and colors indicate the ion used for the [M/H] values, where the vertical error bars are the error on the [M/H] measurements, and our four new $z > 4.7$ DLAs are additionally outlined in black. Note that the larger error bars on the [Fe/H] values are driven by the error propagation from our $\alpha$-enhancement and dust depletion correction method. The upper and lower limits for J2207-0146 at $z = 4.722$ are indicated by black up and down arrows. Although these limits are not included in further analysis, we plot them here to demonstrate how this DLA would lie in our total sample, should it be better constrained.}
    \label{fig: sample ions}
\end{figure*}

We update the R12 and R14 DLA samples from 258 DLAs to 306 DLAs. Within this total sample, we increase the number of DLAs at $z > 4.7$ from 14 to 18 DLAs. We plot the metallicities as a function of redshift for the new full DLA sample in the left panel of Fig. \ref{fig: sample ions}. Our four new high-redshift DLA metallicities are plotted as red stars, the new DLAs we have included from the last decade of publications are plotted as blue squares, and the previously reported DLA sample is plotted in the background in yellow circles. The right panel shows the ions used to measure the metallicity for each DLA and their individual uncertainties. We follow the hierarchy of ion preference described in R12 when considering which metallicity to include in our sample calculations (in order of preference: O, S, Si, Zn, Fe, Ni, Al). The four new high-redshift DLAs are highlighted with a black border to emphasize where they fall in our full sample and which ions were used for their reported metallicities.

We calculate an updated redshift-metallicity relation of DLAs, where we use the new larger sample of 306 DLAs. We bin our $z < 4.7$ DLAs into ten roughly equal bins, where most bins have 29 DLAs/bin except for three bins which have 27 ($3.22 < z < 3.431$), 28 ($2.37 < z < 2.625$), and 30 ($2.813 < z < 3.22$) DLAs each. For each of these bins, we compute a cosmological mean metallicity value for the neutral hydrogen gas as follows:
\begin{equation}
    \langle Z \rangle = {\rm log} \left[\frac{  \sum_{i} 10^{{[M/H]}^i}\mbox{\NHI}^i}{\sum_{i} \mbox{\NHI}^i} \right],
    \label{eq: Z}
\end{equation}
where {\it i} represents each DLA within the respective bin. 

\begin{figure*}
    \centering
    \includegraphics[width = 7.1 in]{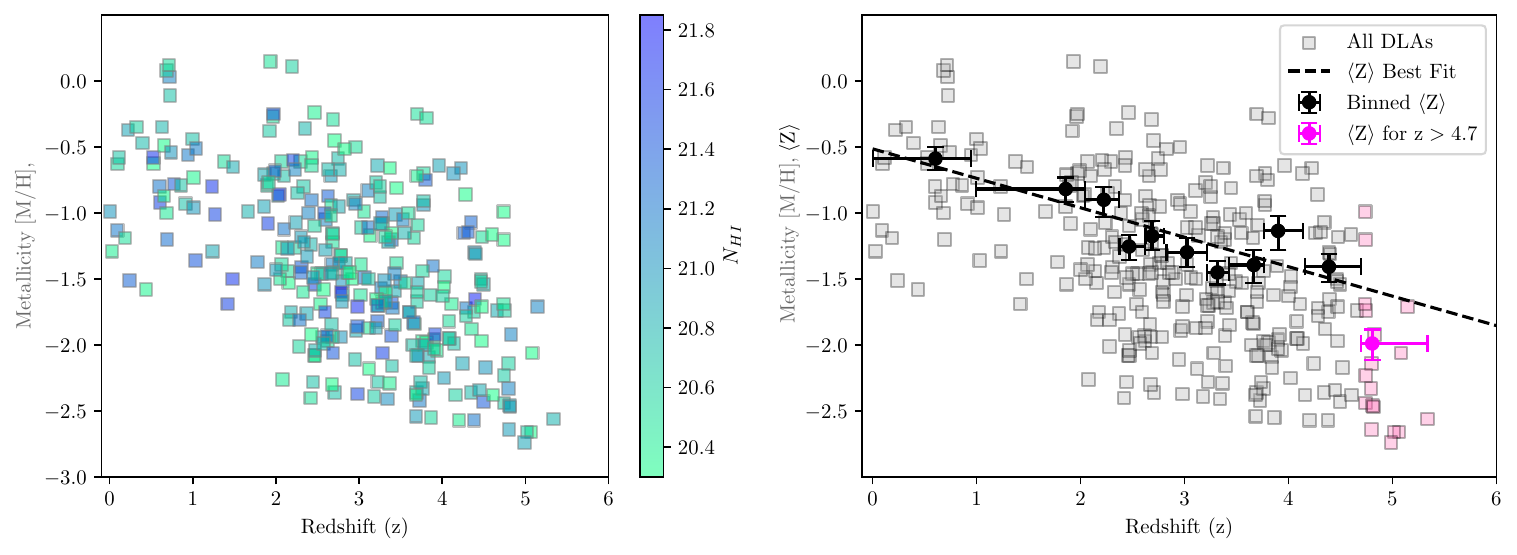} 
    \caption{\emph{Left:} DLA metallicities as a function of redshift, colored by their \NHI value. \emph{Right:} Metallicity evolution of the Universe as a function of redshift. The entirety of our updated 306 DLA sample is plotted in the background in light gray squares, corresponding with the y-axis label for [M/H]. The $\langle Z \rangle$ values for each bin are shown in black for $z < 4.7$ bins and magenta for the $z > 4.7$ bin, corresponding with the black y-axis label for $\langle Z \rangle$. The horizontal error bars show the redshift coverage of each bin, and the vertical uncertainties on the $\langle Z \rangle$ values are calculated using bootstrap routines. A linear regression is fit to the black points and plotted as a black dashed line.}
    \label{fig:metal z trend}
\end{figure*}

We plot the metallicity of the full sample of 306 DLAs as a function of redshift on the right in Fig. \ref{fig:metal z trend}. The $\langle Z \rangle$ values for each of the ten $ z < 4.7$ bins are shown in black, where the x-error bars show the redshift range covered by the respective bin and the y-error is a bootstrap error on the $\langle Z \rangle$ statistic. A linear regression is fit to the $z < 4.7$ binned $\langle Z \rangle$ values and over-plotted as a black dashed line. The magenta point is the same global metallicity statistic ($\langle Z \rangle$), but calculated for the 18 DLAs at $z > 4.7$. The DLAs that make up this final bin are intentionally withheld from the linear regression fit because we will investigate whether their metallicities deviate from the trend at lower redshifts. The left panel of Fig. \ref{fig:metal z trend} shows the \NHI values for all DLAs we include in our sample. 

The uncertainties for all $\langle Z \rangle$ points are calculated using a {\tt SciPy} percentile bootstrap routine with 10,000 resamples of paired combinations of each DLA \NHI\ and [M/H] values within each redshift bin. We use the upper and lower confidence interval of the 68th percentile bootstrap distribution as the uncertainty on the $\langle Z \rangle$ statistic for each redshift bin. 

We derive a linear fit 
\begin{equation}
\langle Z \rangle = (-0.22 \pm 0.05) \, z - (0.52 \pm 0.14)
\label{eqn:Z}
\end{equation}
from the binned evaluations at $z<4.7$. The slope of this trend provides a description of the evolution of the metal content of the neutral gas in the Universe over approximately 12.3 Gyr. Previously R14 had fewer DLAs in the low redshift regime, where the global $\langle Z \rangle$ trend is more sensitive to the intrinsic scatter of the distribution due to small sample size. The addition of the four lowest redshift blue squares as well as the new and updated DLA metallicities from L13 into our lowest redshift bins provides a more robust measure of the global metallicity content of the Universe at these redshifts. 

Of the 18 total DLAs in the $z > 4.7$ range, 16 fall below the linearly extrapolated value of the trend at the median redshift of the $z > 4.7$ bin. If this high-redshift regime followed the linear trend, we would expect  $\langle Z \rangle = -1.59$ at $z = 4.81$. However, comparing the sample of individual DLA metallicity measurements at this redshift to the predicted value based on the $ z < 4.7$ metallicity values yields a $4.4\sigma$ deviation from the prediction based on the uncertainty of $\langle Z \rangle$. The column density weighted cosmic metallicity value itself (the pink point) is $\langle Z \rangle = -2.00^{+0.10}_{-0.13}$ with the sulfur dust depletion and $\alpha$-enhancement correction, which lies 3.8$\sigma$ from the value expected from extrapolating the trend to $ z = 4.81$. Therefore, we observe that both the column density weighted average and the population of individual DLAs at $z > 4.7$ are deviant from the expected trend. Note that if zinc is used to apply the dust depletion and $\alpha$-enhancement correction instead of sulfur like we do for this analysis, the highest redshift bin also has $\langle Z \rangle = -2.00^{+0.10}_{-0.12}$ (see Appendix \ref{sec: appendix_a}), which is well within the error of the $\langle Z \rangle$ value we obtain using the sulfur depletion correction method. We note the difference between our bootstrapping method and the method used in R14 in Section \ref{sec: sig}, as the alternative method we use contributes to the differences in deviation of the high-redshift population.

\begin{figure}
    \centering
    \includegraphics[width = 3.4 in]{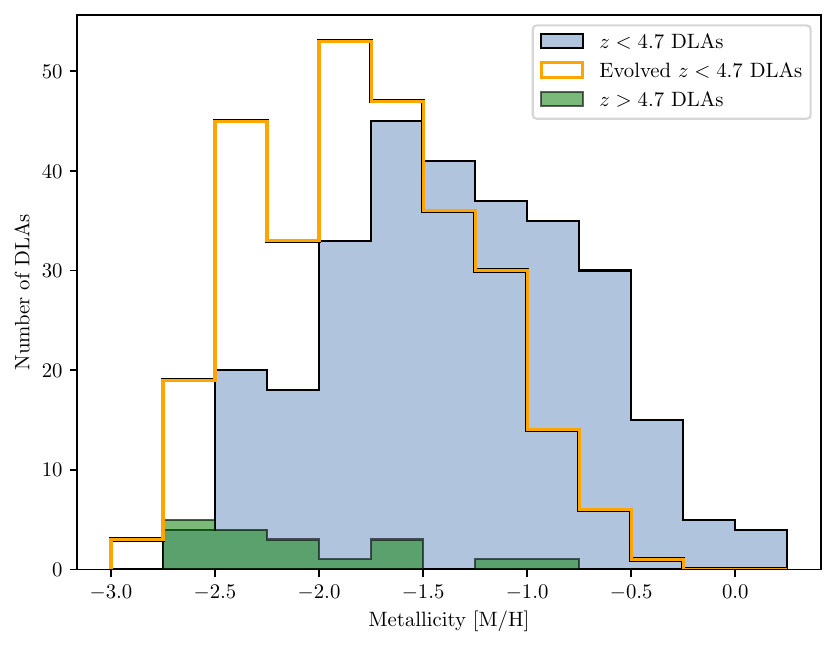}
    \caption{Histogram of DLA metallicity distributions for the $z < 4.7$ sample (blue), the $z > 4.7$ sample (green), and the $z < 4.7$ sample evolved to $ z = 4.81$ (orange). A comparison of the orange and green histograms show the $2.4\sigma$ difference in distributions of metallicity values.}
    \label{fig:metal comp hist}
\end{figure}

Moreover, we may compare the unweighted distribution of metallicity values between the $z<4.7$ and highest redshift bin as follows.  We construct a ``redshift-evolved'' distribution of metallicities from the $z<4.7$ measurements by applying Equation~\ref{eqn:Z} for $z=4.81$ for each. We compare this evolved distribution of metallicities to the distribution of metallicities at $z > 4.7$ in Fig. \ref{fig:metal comp hist}. The original metallicity distribution for DLAs at $z < 4.7$ is in blue and the evolved distribution is outlined in orange. We use a two-sided Kolmogorov-Smirnov (K-S) test diagnostic to compare the orange evolved histogram with the $z > 4.7$ distribution of metallicity values, shown in green. The K-S test yields that we can reject the hypothesis that the two samples are drawn from the same parent sample with $2.4\sigma$ confidence. 

Although we do not include the DLA at $ z = 4.772$ towards J2207-0146 in this high-redshift sample as we only have limits on its metallicity, we do test its impact on the trend and significance outcomes. We test including this DLA in our sample by using the upper and lower limits as well as an appropriate intermediate value of $\rm [M/H] = -0.6$, based on the average of the upper and lower limits reported in Section \ref{sec: J0027 @ z = 4.722}. Since we can only measure limits for this target, we do not include it in our final sample of DLAs unbiased in metallicity. The overall global metallicity trend does not change significantly when this DLA is included due to its lower $N_{\rm HI}$ column density within the DLA range, but the $\sigma$ deviation of the high-z bin (pink point in Fig. \ref{fig:metal z trend} from the linear $\langle Z \rangle$ fit does change depending on if the true metallicity lies closer to the upper or lower limit. For the lower limit, the deviation of the high-redshift $\langle \rm Z \rangle$ becomes $\sigma = 4.07$, which is consistent within the error. For the best intermediate value of $\rm [M/H] = -0.6 ~\pm~ 0.3$, the deviation becomes $\sigma = 3.84$. Finally, for the upper limit, the deviation becomes $\sigma = 3.69$.

Finally, if we relax our proximity threshold to $ < 3000 \rm km/s$, the global mean metallicity trend becomes $\langle Z \rangle = (-0.22 \pm 0.04) \, z - (0.53 \pm 0.12)$ with a 4.6$\sigma$ deviation of the $ z > 4.7$ bin from the trend predicted value at the redshift of that bin. Thus, the relaxed limit is in agreement with our results using the more conservative threshold. 

\section{Discussion}
\label{sec:discussion}
In this section we discuss previously reported methods for account for the depletion of metals onto dust grains as well as possible future works to improve these corrections. We also address the significance of our results, how they compare to previous work, and propose physical interpretations. We finally investigate our observations of the behavior of the high-redshift DLA sample in relation to expectations from modeling and simulations of  metal content evolution and CGM composition. 

\subsection{Dust Depletion and $\alpha$-Enhancement Corrections}
We have introduced a new method that corrects a given observed elemental abundance on the basis of $\alpha$-enhancement and dust depletion factors. Many of the current methods for correcting for dust depletion of metals involve using high resolution observations of multiple lines. However, at higher redshifts, we are limited by the nature of the observations -- typically high redshift DLAs are observed for a shorter wavelength range so there are fewer line transitions available for analysis per target. For example, we have shown in Figs. \ref{fig:J0025} - \ref{fig:J2207} all the available lines for the five new DLAs we analyze in this work. There are fewer than three usable line transitions for each target (if any), and the corrections necessary for any given individual DLA are highly uncertain. Given that we are limited in this way, we conclude that the method we present in this work of measuring the average $\alpha$-enhancement and dust depletion correction for a sample at a given metallicity is a solid methodology and better than the alternative of not applying a depletion correction. While the uncertainty of this methodology is large, we add the uncertainty from the correction in quadrature in the reported uncertainty. We test the impact of the inclusion of non-$\alpha$-elements and the higher depletion of the $z < 1.5$ population of DLAs on the overall linear trend and recovered deviation of the $z > 4.7$ population independently in Appendix \ref{sec: appendix_b}. We also find that the resulting metallicity evolution (slope) is consistent with the main result presented including the full sample with $\alpha$-element and dust depletion corrections.

\subsection{Significance}
\label{sec: sig}
We conduct two diagnostics to compare our high-redshift sample to our  $z < 4.7$ sample. The first is the deviation of $\langle Z \rangle$ at $z > 4.7$  from the linear trend fit to $\langle Z \rangle$  at $z < 4.7$ (Equation~\ref{eqn:Z}). The second is the  K-S test result from comparing the metallicity distributions of the evolved $z < 4.7$ sample and the $z > 4.7$ sample. These independent tests yield a $3-5\sigma$ confidence result that the metallicities of high-redshift DLAs deviate from those observed at $z < 4.7$ . 

We report the magnitude of the  $\langle Z \rangle$ value at $z > 4.7$ to be  $\langle Z \rangle = -2.00^{+0.10}_{-0.13}$ with the sulfur dust depletion and $\alpha$-enhancement correction,
$\langle Z \rangle = -2.00^{+0.10}_{-0.12}$ with the zinc dust depletion and $\alpha$-enhancement correction applied, and $\langle Z \rangle = -2.03^{+0.10}_{-0.11}$ with only the $\alpha$-enhancement correction applied. The measurement of these three values are consistent well within the error of each other. 

Additionally, all three values with and without dust depletion and $\alpha$-enhancement corrections applied  are consistent with the  $\langle Z \rangle = -2.03^{+0.09}_{-0.11}$ value reported from R14, within the error. Our high-redshift population deviation from the trend prediction is also consistent with the $~3\sigma$ K-S test result from R14. However, R14 also quantified the significance of the metallicity deviation of the high-redshift sample through a bootstrapping method, which resulted in a significance of $6\sigma$. This value is higher than what their K-S test yielded and higher than our measured significance of $\sim 4.5\sigma$. We associate this difference to the method of bootstrapping employed for estimating the vertical error on the $\langle Z \rangle$ calculations. Their method replaces 1/e of the sample per iteration, assuming the average bootstrap instance for each. Instead, our method replaces the full sample which on average duplicates 1/e of the sample per iteration. The result is that our method has a larger variance that more accurately describes the underlying uncertainty. Randomly matching metallicity values with alternative \HI\ column densities (as in R14) probes the distribution of individual DLAs in each bin, rather than the metallicity evolution of metals in the Universe as a whole. Thus, R14 recover a larger significance than what we measure, despite the samples being relatively similar.   

\subsection{Comparisons with Simulations}
\label{sec: sims}
 Many properties of high-redshift DLAs are elusive due to the distinct observational challenges of the high redshift regime ($z > 4$).Thus, it can be interesting to turn to simulations to enhance our understanding of the possibilities. There are various types of simulations that constrain observed properties of DLAs at $z \sim 2-3$, which is also the most well constrained observational regime of DLAs. For example, the GAlaxy Evolution and Assembly (GAEA) semi-analytic models estimate \HI\ and gas properties in DLAs out to z = 4, and are consistent with observations at $z \sim 2-3$, however their model breaks down at higher redshifts. Even at $z \sim 2-3$, they observe an excess of low-metallicity systems that are not found observationally \citep{DiGioia2020}. At either extreme of the $z \sim 2-3$ range, DLAs are much harder to constrain both observationally and in simulations and modeling. There have been recent attempts to incorporate DLAs in $z < 2$ simulations \citep{Garratt-Smithson2021}, however models are still not well matched with existing observations. On the other end, there are even fewer simulations that are able to recreate any sort of DLA conditions at $z > 4$. \citet{Nagamine2004} presents a series of smoothed particle hydrodynamics simulations over $z = 0-4.5$ that focus on the distribution of SFR and metallicities in DLAs at those redshifts. The metallicity evolution produced is consistent with observations however the mean metallicities are much higher than observed values at all redshifts.  

There are a handful of simulations that are able to predict specific components of the observed redshift metallicity relation we present in this work. N-body simulations from \citet{Maio2013} predict an average mean metallicity of $\sim 10^{-3} Z_{\odot}$ at $z \sim 7$, consistent with the observed metallicity floor. Additionally, adaptive mesh refinement simulations such as the one presented in \citet{Wise2012} are also able to predict the observed metallicity floor of DLAs first reported in R12 via the transition from Population III stars to Population II stars through pair-instability supernovae. The n512RT64 from the TECHNICOLOR DAWN simulation suite \citep{Finlator2018} is a cosmological radiation hydrodynamical simulation that predicts a HI-weighted mean metallicity of $\rm [M/H] = -1.72$ for DLAs at z = 5. This average value is higher than the $\langle Z \rangle$ value that we measure for our $z > 4.7$ bin within 2$\sigma$, and is thus more consistent with the expected value from the linear trend fit to the $z < 4.7$ $\langle Z \rangle$ values. Thus, there is some tension between theoretical and model expectations of the metallicity-redshift relationship for DLAs and the metal content of the Universe at high redshifts. Even though some simulations are coming close to matching observed properties, there is tension between results due to the complex nature of the baryon cycle and gas flows in galaxies towards the beginning of the Universe. Many simulations ignore the impacts of stellar feedback winds and assume an instantaneous process of recycling of gas and metals from the life cycle of stars-- both processes that are pertinent to how DLAs can be stripped of metals or become more metal rich \citep{Hirschmann2016}. Future work on clarifying feedback cycles in galaxy simulations and theoretical predictions will hopefully start to relieve some of these existing tensions.

\subsection{Interpretation of Results}
Finally, our results can be used to consider explanations for the sudden change in the metallicity evolution of cosmic gas at $z\sim 5$. Originally, \citet{Rafelski2014} proposed that this decrease was driven by differences in the gas associated with DLAs: while DLAs probe the CGM of galaxies at $z = 0-4$, they may probe gas in the outer CGM and/or beyond the viral radius at $z\sim 5$, where the gas is less enriched. This interpretation still explains the data, especially in light of recent observations suggesting that the impact parameters of DLA galaxies might increase with redshift \citep{Kanekar2018, kanekar2020, neeleman2017, neeleman2018, neeleman2025, oyarzun2024}. However, and as also pointed out by R14, this drop in metallicity at $z > 4.7$ cannot yet be ruled out as a statistical Type I error, i.e., that the drop in metallicity is driven by the observation of 20 particularly metal-poor, although unbiased, DLAs. 

One may also consider that a sharp decrease in metallicity at $z > 4.7$ is consistent with galaxy formation being non-linear with redshift. In this regard, there is a need for galaxy formation models to study the metallicity-redshift evolution in greater detail, with emphasis on qualitatively and quantitatively constraining the physical processes driving its shape. Particularly interesting would be to determine if there is any connection between the sharp decrease in metallicity and the end of the reionization epoch, especially given their proximity in cosmic history \citep[e.g.][]{robertson2015, becker2024, Bosman2024, roth2023, zhu2022, zhu2023}.

\section{Summary}
\label{sec: summary}
We present a sample of DLAs unbiased in metallicity with  metallicity measurements out to $z\sim 5$. In addition to 62 DLAs from the past decade of literature, we report four recently discovered DLAs at $z > 4.7$ DLAs observed with high-resolution spectroscopy from X-shooter and HIRES. We use this new sample to revisit the redshift-metallicity trend for DLAs out to $z\sim 5$ characterized in previous work (\citealt{Rafelski2012, Rafelski2014}).

After binning our DLAs according to their redshift, we calculate the average metallicity of the Universe (i.e. $\langle Z \rangle$, Eq. \ref{eq: Z}). A linear trend fit to the $z < 4.7$ redshift bins yields $\langle Z \rangle = (-0.22 \pm 0.04)\,z - (0.51 \pm 0.13)$. We find that $\langle Z \rangle$ at $z > 4.7$ falls 4.4$\sigma$ below the value expected at these redshifts per the global metallicity trend fit. A K-S test rejects the probability that the $z > 4.7$ sample and an evolved $z < 4.7$ sample are equivalent with over 3$\sigma$ confidence. Thus, our updated sample of DLAs unbiased in metallicity upholds the drastic turnover observed in R14 at $z \gtrsim 4.7$ and we conclude that this turnover in the metal content of the Universe is occurring near $z \gtrsim 4.7$ at high confidence.  
In summary, we:
\begin{itemize}
    \item Add 4 new DLA metallicity measurements at $ z > 4.7$.
    \item Compile 62 DLA metallicities reported in the literature in the last decade, many of which are at intermediate redshift bins and therefore improve the $\langle Z \rangle$ measurement.
    \item Introduce a novel dust-depletion correction routine which primarily impacts the lowest redshift bins and asserts minimal to no corrections required for the majority of the $z > 4.7$ DLA metallicities.
    \item Re-analyze the $z > 4.7$ population of DLAs in comparison to the results from R12 and R14 to better characterize the evolution of cosmic metallicity.
\end{itemize}
With the advent of the Dark Energy Spectrograph Instrument (DESI) survey, discoveries from Euclid, and other large upcoming surveys, larger samples of $z > 5$ QSOs will be observed and followed up with high-resolution spectroscopy, allowing for the discovery of more high-redshift DLAs and increased sample sizes. 

\startlongtable
\begin{deluxetable*}{lccccccc}
\tabletypesize{\normalsize}
\label{tab:sample}

\tablecaption{Full sample of 306 DLAs. The targets listed in bold are our VLT/X-shooter DLA sample properties with updated redshifts calculated from HIRES observations. The values in the $\rm [X/H]_{obs}$ column are the observed abundance values as reported in the referenced literature source with no $\alpha-$enhancement correction applied. The values in the $\rm[M/H]_{corr}^S$ column are the new metallicities with the dust depletion and $\alpha-$enhancement correction applied, as measured using the sulfur relationship in Fig. \ref{fig:s_fe}.\\
${^a}$ F17:\citet{Fynbo2017}; H18: \citet{Heintz2018}; S16: \citet{Srianand2016}; B16: \citet{Berg2016}; K18: \citet{Kanekar2018}; B19: \citet{Bashir2019}; O14: \citet{Oliveira2014}; K14: \citet{Kanekar2014}; H25: \citet{Huyan2025}; F14: \citet{Fumagalli2014}; R14: \citet{Rafelski2014}; L13:\citet{Lehner2013}; R12: \citet{Rafelski2012}}
\setlength{\tabcolsep}{0.11in}

\tablehead{QSO & $z_{QSO}$ & $z_{abs}$ & $\log\mbox{\NHI}$ & $\rm [X/H]_{obs}$ &$\rm[M/H]_{corr}^S$ & ${f_{mtl}}$ & ${\rm Ref.^a}$}
\startdata
\bf{J2207-0146} & 5.53 & 5.337 & 20.80~$\pm$~0.20 & -2.56~$\pm$~0.22 & -2.56~$\pm$~0.22  & Si & This Work \\
\bf{J0306+1853} & 5.36 & 4.987 & 20.90~$\pm$~0.20 & -3.01~$\pm$~0.20 & -2.74~$\pm$~0.31  & Fe &This Work \\
J1437+2323 & 5.32 & 4.801 & 20.95~$\pm$~0.15 & -2.64~$\pm$~0.15 & -2.64~$\pm$~0.15  & Si & R14 \\
J1202+3235 & 5.292 & 4.795 & 21.10~$\pm$~0.15 & -2.60~$\pm$~0.16 & -2.33~$\pm$~0.29  & Fe &R12 \\
J1202+3235 & 5.292 & 5.065 & 20.30~$\pm$~0.15 & -2.66~$\pm$~0.16 & -2.66~$\pm$~0.16  & Si & R12 \\
J1208+0010 & 5.27 & 5.082 & 20.30~$\pm$~0.15 & -2.06~$\pm$~0.15 & -2.06~$\pm$~0.15  & Si & R14 \\
\bf{J0747+1153} & 5.26 & 5.145 & 21.10~$\pm$~0.20 & -1.71~$\pm$~0.20 & -1.71~$\pm$~0.20  & S &This Work \\
J1221+4445 & 5.206 & 4.811 & 20.90~$\pm$~0.20 & -2.46~$\pm$~0.20 & -2.46~$\pm$~0.20  & Si & R14 \\
J0824+1302 & 5.188 & 4.472 & 20.30~$\pm$~0.10 & -1.97~$\pm$~0.12 & -1.97~$\pm$~0.12  & Si & R14 \\
J1054+1633 & 5.187 & 4.817 & 20.95~$\pm$~0.15 & -2.47~$\pm$~0.15 & -2.47~$\pm$~0.15  & Si & R14 \\
J1054+1633 & 5.187 & 4.135 & 21.00~$\pm$~0.10 & -0.70~$\pm$~0.11 & -0.70~$\pm$~0.11  & Si & R14 \\
J1054+1633 & 5.187 & 3.842 & 20.60~$\pm$~0.20 & -2.17~$\pm$~0.18 & -2.17~$\pm$~0.18  & Si & R14 \\
J1626+2751 & 5.1847 & 4.311 & 21.34~$\pm$~0.15 & -1.58~$\pm$~0.17 & -1.17~$\pm$~0.29  & Fe &R12 \\
J1626+2751 & 5.1847 & 4.497 & 21.39~$\pm$~0.15 & -2.70~$\pm$~0.16 & -2.43~$\pm$~0.29  & Fe &R12 \\
J1132+1209&5.167&5.016&20.65~$\pm$~0.20&-2.66~$\pm$~0.20&2.66~$\pm$~0.20&Si&R14\\
J1132+1209 & 5.167 & 4.38 & 21.20~$\pm$~0.20 & -2.57~$\pm$~0.17 & -2.57~$\pm$~0.17  & Si & R14 \\
J1204-0021 & 5.09 & 3.644 & 20.70~$\pm$~0.10 & -2.00~$\pm$~0.17 & -2.00~$\pm$~0.17  & Si & R14 \\
\bf{J0025-0145} & 5.07 & 4.739 & 20.30~$\pm$~0.20 & -1.41~$\pm$~0.20 &-1.00~$\pm$~0.31  & Fe &This Work\\
J1013+4240 & 5.038 & 4.798 & 20.60~$\pm$~0.15 & -2.14~$\pm$~0.15 & -2.14~$\pm$~0.15  & Si & R12 \\
J1340+3926 & 5.026 & 4.826 & 21.10~$\pm$~0.10 & -1.92~$\pm$~0.17 & -1.92~$\pm$~0.17  & Si & R14 \\
J1626+2858 & 5.022 & 4.608 & 20.30~$\pm$~0.15 & -2.38~$\pm$~0.18 & -2.38~$\pm$~0.18  & Si & R14 \\
J1101+0531 & 4.986 & 4.345 & 21.30~$\pm$~0.10 & -1.07~$\pm$~0.12 & -1.07~$\pm$~0.12  & Si & R12 \\
J0040-0915 & 4.976 & 4.739 & 20.30~$\pm$~0.15 & -1.64~$\pm$~0.17 & -1.23~$\pm$~0.29  & Fe & R12 \\
J1737+5828 & 4.941 & 4.744 & 20.65~$\pm$~0.20 & -2.23~$\pm$~0.21 & -2.23~$\pm$~0.21  & Si & R14 \\
SDSS1737+5828 & 4.941 & 4.743 & 20.65~$\pm$~0.15 & -2.71~$\pm$~0.22 & -2.44~$\pm$~0.33  & Fe &R12 \\
J1245+3822 & 4.94 & 4.447 & 20.85~$\pm$~0.10 & -2.34~$\pm$~0.11 & -2.34~$\pm$~0.11  & Si & R14 \\
J1253+1046 & 4.908 & 4.6 & 20.30~$\pm$~0.15 & -1.60~$\pm$~0.16 & -1.19~$\pm$~0.28  & Fe &R12 \\
J2252+1425 & 4.906 & 4.747 & 20.60~$\pm$~0.15 & -2.01~$\pm$~0.19 & -1.74~$\pm$~0.31  & Fe &R12 \\
J1051+3545 & 4.899 & 4.35 & 20.45~$\pm$~0.10 & -1.88~$\pm$~0.10 & -1.88~$\pm$~0.10  & Si & R12 \\
J0831+4046 & 4.885 & 4.344 & 20.75~$\pm$~0.15 & -2.36~$\pm$~0.15 & -2.36~$\pm$~0.15  & Si & R12 \\
J1129-1526 & 4.87 & 4.495 & 20.80~$\pm$~0.20 & -1.54~$\pm$~0.14 & -1.54~$\pm$~0.14  & Si & H25 \\
J0759+1800 & 4.862 & 4.658 & 20.85~$\pm$~0.15 & -1.74~$\pm$~0.16 & -1.74~$\pm$~0.16  & Si & R12 \\
J2328+0217 & 4.86 & 4.737 & 20.48~$\pm$~0.20 & -1.69~$\pm$~0.11 & -1.69~$\pm$~0.11  & Si & H25 \\
J1418+3142 & 4.85 & 3.962 & 20.90~$\pm$~0.15 & -0.64~$\pm$~0.15 & -0.64~$\pm$~0.15  & Si & R14 \\
J1607+1604 & 4.798 & 4.474 & 20.30~$\pm$~0.15 & -1.71~$\pm$~0.15 & -1.71~$\pm$~0.15  & Si & R12 \\
J0307-4945 & 4.78 & 4.468 & 20.67~$\pm$~0.09 & -1.50~$\pm$~0.12 & -1.5~$\pm$~0.12  & Si & R12 \\
J0839+3524 & 4.777 & 4.28 & 20.30~$\pm$~0.15 & -1.39~$\pm$~0.17 & -0.87~$\pm$~0.29  & Fe &R12 \\
J1336-1830 & 4.75 & 4.519 & 21.00~$\pm$~0.20 & -2.17~$\pm$~0.14 & -2.17~$\pm$~0.14  & Si & H25 \\
J1336-1830 & 4.75 & 4.286 & 20.60~$\pm$~0.20 & -1.78~$\pm$~0.13 & -1.78~$\pm$~0.13  & Si &H25 \\
J1200+4618 & 4.73 & 4.476 & 20.50~$\pm$~0.15 & -1.62~$\pm$~0.16 & -1.21~$\pm$~0.28  & Fe &R12 \\
J0307-4945 & 4.716 & 3.591 & 20.5~$\pm$~0.15 & -1.78~$\pm$~0.15 & -1.00~$\pm$~0.28  & Fe &B16 \\
J0307-4945 & 4.716 & 4.466 & 20.6~$\pm$~0.1 & -1.52~$\pm$~0.1 & -1.52~$\pm$~0.1 & Si& B16 \\
J1100+1122 & 4.707 & 4.395 & 21.74~$\pm$~0.10 & -1.92~$\pm$~0.18 & -1.65~$\pm$~0.30  & Fe &R12 \\
J235152.80+160048 & 4.694 & 3.786 & 20.85~$\pm$~0.10 & -2.33~$\pm$~0.20 & -1.99~$\pm$~0.31  & Fe &F14 \\
BR1202-07 & 4.69 & 4.383 & 20.60~$\pm$~0.14 & -1.75~$\pm$~0.14 & -1.75~$\pm$~0.14 & Si& R12 \\
J1042+3107 & 4.688 & 4.087 & 20.75~$\pm$~0.10 & -1.95~$\pm$~0.10 & -1.95~$\pm$~0.10 & Si& R12 \\
J1654+2227 & 4.678 & 4.002 & 20.60~$\pm$~0.15 & -1.90~$\pm$~0.16 & -1.63~$\pm$~0.29  & Fe &R12 \\
J1438+4314 & 4.611 & 4.399 & 20.89~$\pm$~0.15 & -1.31~$\pm$~0.15 & -1.31~$\pm$~0.15 & Si& R12 \\
J1201+2117 & 4.579 & 3.797 & 21.35~$\pm$~0.15 & -0.75~$\pm$~0.15 & -0.75~$\pm$~0.15 & Si& R12 \\
J1201+2117 & 4.579 & 4.158 & 20.60~$\pm$~0.15 & -2.38~$\pm$~0.15 & -2.38~$\pm$~0.15 & Si& R12 \\
BR2237-0607 & 4.56 & 4.08 & 20.52~$\pm$~0.11 & -1.82~$\pm$~0.11 & -1.82~$\pm$~0.11 & Si& R12 \\
J2239-0552 & 4.557 & 4.08 & 20.6~$\pm$~0.1 & -1.95~$\pm$~0.1 & -1.95~$\pm$~0.1 & Si& B16 \\
BR0019-15 & 4.53 & 3.439 & 20.92~$\pm$~0.10 & -1.01~$\pm$~0.11 & -1.01~$\pm$~0.11 & Si& R12 \\
PSS0134+3317 & 4.52 & 3.761 & 20.85~$\pm$~0.05 & -2.60~$\pm$~0.16 & -2.33~$\pm$~0.29  & Fe &R12 \\
PSS1715+3809 & 4.52 & 3.341 & 21.05~$\pm$~0.15 & -2.68~$\pm$~0.17 & -2.41~$\pm$~0.29  & Fe &R12 \\
PSS1723+2243 & 4.52 & 3.695 & 20.50~$\pm$~0.15 & -0.82~$\pm$~0.20 & -0.24~$\pm$~0.31  & Fe &R12 \\
PSS2315+0921 & 4.52 & 3.425 & 21.10~$\pm$~0.20 & -1.46~$\pm$~0.21 & -1.46~$\pm$~0.21 & Si& R12 \\
J085143.72+233208 & 4.499 & 3.53 & 21.10~$\pm$~0.10 & -1.05~$\pm$~0.15 & -1.05~$\pm$~0.15  & Zn &F14 \\
J0834+2140 & 4.497 & 3.71 & 20.85~$\pm$~0.10 & -1.80~$\pm$~0.16 & -1.53~$\pm$~0.29  & Fe &R12 \\
J0834+2140 & 4.497 & 4.39 & 21.00~$\pm$~0.20 & -1.30~$\pm$~0.20 & -1.30~$\pm$~0.20 & Si& R12 \\
PC0953+47 & 4.46 & 3.891 & 21.20~$\pm$~0.10 & -1.67~$\pm$~0.15 & -1.40~$\pm$~0.28  & Fe &R12 \\
PC0953+47 & 4.46 & 3.403 & 21.15~$\pm$~0.15 & -1.99~$\pm$~0.29 & -1.72~$\pm$~0.38  & Fe &R12 \\
PC0953+47 & 4.46 & 4.244 & 20.90~$\pm$~0.15 & -2.14~$\pm$~0.15 & -2.14~$\pm$~0.15 & Si& R12 \\
PSS0808+52 & 4.45 & 3.113 & 20.65~$\pm$~0.07 & -1.50~$\pm$~0.14 & -1.50~$\pm$~0.14 & Si& R12 \\
PSS2241+1352 & 4.44 & 4.283 & 21.15~$\pm$~0.10 & -1.72~$\pm$~0.11 & -1.72~$\pm$~0.11 & Si& R12 \\
J0006-6208 & 4.44 & 3.203 & 20.9~$\pm$~0.15 & -1.90~$\pm$~0.15 & -1.87~$\pm$~0.28  & Fe &B16 \\
J0006-6208 & 4.44 & 3.775 & 21~$\pm$~0.2 & -0.94~$\pm$~0.2 & -0.94~$\pm$~0.2  & Zn &B16 \\
J0747+4434 & 4.432 & 4.02 & 20.95~$\pm$~0.15 & -2.52~$\pm$~0.20 & -2.25~$\pm$~0.31  & Fe &R12 \\
BRI0952-01 & 4.43 & 4.024 & 20.55~$\pm$~0.10 & -1.75~$\pm$~0.18 & -1.48~$\pm$~0.30  & Fe &R12 \\
J1412+0624 & 4.421 & 4.109 & 20.40~$\pm$~0.15 & -1.96~$\pm$~0.18 & -1.69~$\pm$~0.3  & Fe &R12 \\
PSS1443+27 & 4.41 & 4.224 & 21.00~$\pm$~0.10 & -1.19~$\pm$~0.17 & -0.67~$\pm$~0.29  & Fe &R12 \\
BR1013+0035 & 4.405 & 3.104 & 21.10~$\pm$~0.10 & -0.83~$\pm$~0.10 & -0.83~$\pm$~0.10 & Si& R12 \\
J0817+1351 & 4.389 & 4.258 & 21.30~$\pm$~0.15 & -1.15~$\pm$~0.15 & -1.15~$\pm$~0.15 & Si& R12 \\
BR0951-04 & 4.37 & 3.857 & 20.60~$\pm$~0.10 & -1.89~$\pm$~0.17 & -1.62~$\pm$~0.29  & Fe &R12 \\
BR0951-04 & 4.37 & 4.203 & 20.40~$\pm$~0.10 & -2.57~$\pm$~0.10 & -2.57~$\pm$~0.10 & Si& R12 \\
J1633+1411 & 4.365 & 2.882 & 20.3~$\pm$~0.15 & -2.08~$\pm$~0.15 & -1.45~$\pm$~0.28  & Fe &B16 \\
J1248+3110 & 4.355 & 3.697 & 20.60~$\pm$~0.10 & -1.87~$\pm$~0.16 & -1.60~$\pm$~0.29  & Fe &R12 \\
PSS1248+31 & 4.355 & 3.697 & 20.63~$\pm$~0.07 & -1.67~$\pm$~0.07 & -1.67~$\pm$~0.07  & Si & R12 \\
J1058+1245 & 4.341 & 3.432 & 20.6~$\pm$~0.1 & -1.85~$\pm$~0.1 & -1.85~$\pm$~0.1  & Si & B16 \\
J0424-2209 & 4.329 & 2.982 & 21.4~$\pm$~0.15 & -1.86~$\pm$~0.16 & -1.86~$\pm$~0.16  & Zn &B16 \\
BRJ0426-2202 & 4.32 & 2.982 & 21.50~$\pm$~0.15 & -2.64~$\pm$~0.17 & -2.37~$\pm$~0.29  & Fe &R12 \\
J0113-2803 & 4.314 & 3.106 & 21.2~$\pm$~0.1 & -1.11~$\pm$~0.1 & -1.11~$\pm$~0.1  & Si & B16 \\
J0234-1806 & 4.305 & 3.693 & 20.4~$\pm$~0.15 & -1.01~$\pm$~0.15 & -0.75~$\pm$~0.28  & Fe &B16 \\
PSS1432+39 & 4.28 & 3.272 & 21.25~$\pm$~0.10 & -1.03~$\pm$~0.11 & -1.03~$\pm$~0.11  & Si & R12 \\
PSS2155+1358 & 4.26 & 3.316 & 20.55~$\pm$~0.15 & -1.18~$\pm$~0.17 & -1.18~$\pm$~0.17  & Si & R12 \\
J1051+3107 & 4.253 & 4.139 & 20.70~$\pm$~0.20 & -1.99~$\pm$~0.21 & -1.99~$\pm$~0.21  & Si & R12 \\
PSS0957+33 & 4.25 & 3.28 & 20.45~$\pm$~0.08 & -1.08~$\pm$~0.10 & -1.08~$\pm$~0.10  & Si & R12 \\
J2344+0342 & 4.248 & 3.22 & 21.3~$\pm$~0.1 & -1.7~$\pm$~0.32 & -1.7~$\pm$~0.32  & Zn &B16 \\
PSS2344+0342 & 4.24 & 3.22 & 21.35~$\pm$~0.07 & -2.09~$\pm$~0.26 & -1.82~$\pm$~0.35  & Fe &R12 \\
J0134+0400 & 4.185 & 3.692 & 20.7~$\pm$~0.1 & -2.71~$\pm$~0.1 & -2.08~$\pm$~0.26  & Fe &B16 \\
J0134+0400 & 4.185 & 3.772 & 20.7~$\pm$~0.1 & -0.91~$\pm$~0.1 & -0.91~$\pm$~0.1  & Si & B16 \\
PSS1506+5220 & 4.18 & 3.224 & 20.67~$\pm$~0.07 & -2.28~$\pm$~0.07 & -2.28~$\pm$~0.07  & Si & R12 \\
PSS1802+5616 & 4.18 & 3.811 & 20.35~$\pm$~0.20 & -1.98~$\pm$~0.22 & -1.98~$\pm$~0.22  & Si & R12 \\
PSS1802+5616 & 4.18 & 3.391 & 20.30~$\pm$~0.10 & -1.59~$\pm$~0.15 & -1.18~$\pm$~0.28  & Fe &R12 \\
PSS1802+5616 & 4.18 & 3.554 & 20.50~$\pm$~0.10 & -1.80~$\pm$~0.17 & -1.53~$\pm$~0.29  & Fe &R12 \\
PSS1802+5616 & 4.18 & 3.762 & 20.55~$\pm$~0.15 & -1.71~$\pm$~0.21 & -1.44~$\pm$~0.32  & Fe &R12 \\
PSS2323+2758 & 4.18 & 3.685 & 20.95~$\pm$~0.10 & -2.54~$\pm$~0.10 & -2.54~$\pm$~0.10  & Si & R12 \\
J0529-3552 & 4.172 & 3.684 & 20.4~$\pm$~0.15 & -2.38~$\pm$~0.15 & -2.38~$\pm$~0.15  & Si & B16 \\
PSS0209+05 & 4.17 & 3.666 & 20.45~$\pm$~0.10 & -1.84~$\pm$~0.10 & -1.84~$\pm$~0.10  & Si & R12 \\
PSS0209+05 & 4.17 & 3.864 & 20.55~$\pm$~0.10 & -2.55~$\pm$~0.10 & -2.55~$\pm$~0.10  & Si & R12 \\
J0132+1341 & 4.152 & 3.936 & 20.4~$\pm$~0.15 & -2.04~$\pm$~0.15 & -2.04~$\pm$~0.15  & Si & B16 \\
J0747+2739 & 4.133 & 3.901 & 20.6~$\pm$~0.15 & -2.03~$\pm$~0.15 & -2.03~$\pm$~0.15  & Si & B16 \\
PSS0133+0400 & 4.13 & 3.774 & 20.55~$\pm$~0.10 & -0.70~$\pm$~0.10 & -0.70~$\pm$~0.10  & Si & R12 \\
PSS0133+0400 & 4.13 & 3.692 & 20.70~$\pm$~0.10 & -2.63~$\pm$~0.17 & -2.36~$\pm$~0.29  & Fe &R12 \\
Q1055+46 & 4.13 & 3.317 & 20.34~$\pm$~0.10 & -1.60~$\pm$~0.15 & -1.60~$\pm$~0.15  & Si & R12 \\
J1057+1910&4.128&3.373 & 20.3~$\pm$~0.1 & -1.55~$\pm$~0.11 & -0.67~$\pm$~0.26  & Fe &B16 \\
J0003-2603 & 4.125 & 3.39 & 21.4~$\pm$~0.1 & -1.93~$\pm$~0.11 & -1.93~$\pm$~0.11  & Zn &B16 \\
J1257-0111 & 4.112 & 4.021 & 20.30~$\pm$~0.10 & -1.56~$\pm$~0.10 & -1.56~$\pm$~0.10  & Si & R12 \\
Q0000-2619 & 4.11 & 3.39 & 21.41~$\pm$~0.08 & -1.68~$\pm$~0.20 & -1.68~$\pm$~0.20  & Si & R12 \\
FJ0747+2739 & 4.11 & 3.9 & 20.50~$\pm$~0.10 & -1.96~$\pm$~0.10 & -1.96~$\pm$~0.10  & Si & R12 \\
FJ0747+2739 & 4.11 & 3.424 & 20.85~$\pm$~0.05 & -1.86~$\pm$~0.20 & -1.59~$\pm$~0.31  & Fe &R12 \\
J0415-4357 & 4.073 & 3.808 & 20.5~$\pm$~0.2 & -0.28~$\pm$~0.2 & -0.28~$\pm$~0.2  & Zn &B16 \\
SDSS0127-00 & 4.06 & 3.728 & 21.15~$\pm$~0.10 & -2.42~$\pm$~0.10 & -2.42~$\pm$~0.10  & Si & R12 \\
PSS0007+2417 & 4.05 & 3.704 & 20.55~$\pm$~0.15 & -1.51~$\pm$~0.27 & -1.10~$\pm$~0.36  & Fe &R12 \\
PSS0007+2417 & 4.05 & 3.496 & 21.10~$\pm$~0.10 & -1.52~$\pm$~0.11 & -1.52~$\pm$~0.11  & Si & R12 \\
PSS0007+2417 & 4.05 & 3.838 & 20.85~$\pm$~0.15 & -2.36~$\pm$~0.24 & -2.09~$\pm$~0.34  & Fe &R12 \\
PSS1253-0228 & 4.01 & 2.783 & 21.85~$\pm$~0.20 & -1.87~$\pm$~0.21 & -1.60~$\pm$~0.32  & Fe &R12 \\
J0255+0048 & 4.003 & 3.256 & 20.9~$\pm$~0.1 & -1.08~$\pm$~0.1 & -1.08~$\pm$~0.1  & Si & B16 \\
J0255+0048 & 4.003 & 3.914 & 21.5~$\pm$~0.1 & -1.92~$\pm$~0.1 & -1.92~$\pm$~0.1  & S &B16 \\
J025518.58+004847 & 3.996 & 3.253 & 20.60~$\pm$~0.15 & -0.80~$\pm$~0.11 & -0.80~$\pm$~0.11  & Si & F14 \\
BRI1346-03 & 3.99 & 3.736 & 20.72~$\pm$~0.10 & -2.28~$\pm$~0.10 & -2.28~$\pm$~0.10  & Si & R12 \\
PSS1535+2943 & 3.99 & 3.202 & 20.65~$\pm$~0.15 & -1.30~$\pm$~0.25 & -0.78~$\pm$~0.35  & Fe &R12 \\
PSS1535+2943 & 3.99 & 3.761 & 20.40~$\pm$~0.15 & -1.97~$\pm$~0.16 & -1.97~$\pm$~0.16  & Si & R12 \\
J0255+00 & 3.97 & 3.253 & 20.70~$\pm$~0.10 & -0.88~$\pm$~0.11 & -0.88~$\pm$~0.11  & Si & R12 \\
BR1117-1329 & 3.96 & 3.35 & 20.84~$\pm$~0.10 & -1.22~$\pm$~0.13 & -1.22~$\pm$~0.13  & Si & R12 \\
BRI1108-07 & 3.92 & 3.608 & 20.50~$\pm$~0.10 & -1.75~$\pm$~0.10 & -1.75~$\pm$~0.10  & Si & R12 \\
J0814+5029 & 3.88 & 3.708 & 21.35~$\pm$~0.15 & -2.08~$\pm$~0.15 & -2.08~$\pm$~0.15  & Si & R12 \\
J0825+3544 & 3.846 & 3.207 & 20.30~$\pm$~0.10 & -1.92~$\pm$~0.16 & -1.65~$\pm$~0.29  & Fe &R12 \\
J0825+3544 & 3.846 & 3.657 & 21.25~$\pm$~0.10 & -1.83~$\pm$~0.13 & -1.83~$\pm$~0.13  & Si & R12 \\
J0124+0044 & 3.837 & 2.261 & 20.7~$\pm$~0.15 & -0.85~$\pm$~0.15 & -0.85~$\pm$~0.15  & Si & B16 \\
J094927.88+111518 & 3.824 & 2.758 & 20.85~$\pm$~0.10 & -0.95~$\pm$~0.10 & -0.95~$\pm$~0.10  & Si & F14 \\
J0909+3303 & 3.7835 & 3.658 & 20.55~$\pm$~0.10 & -1.19~$\pm$~0.11 & -1.19~$\pm$~0.11  & Si & R12 \\
J1312+0841 & 3.731 & 2.66 & 20.50~$\pm$~0.1 & -1.80~$\pm$~0.1 & -1.02~$\pm$~0.26  & Fe &B16 \\
J1552+1005 & 3.722 & 3.601 & 21.1~$\pm$~0.1 & -1.75~$\pm$~0.11 & -1.75~$\pm$~0.11  & S &B16 \\
J090810.36+023818 & 3.71 & 2.959 & 21.10~$\pm$~0.10 & -0.93~$\pm$~0.12 & -0.93~$\pm$~0.12  & Si & F14 \\
B1418-064 & 3.689 & 3.448 & 20.50~$\pm$~0.10 & -1.38~$\pm$~0.13 & -1.38~$\pm$~0.13  & Si & R12 \\
J1421-0643 & 3.688 & 3.449 & 20.30~$\pm$~0.15 & -1.70~$\pm$~0.15 & -0.82~$\pm$~0.28  & Fe &B16 \\
J1108+1209 & 3.679 & 3.397 & 20.7~$\pm$~0.1 & -2.79~$\pm$~0.1 & -2.16~$\pm$~0.26  & Fe &B16 \\
J1108+1209 & 3.679 & 3.546 & 20.8~$\pm$~0.15 & -1.15~$\pm$~0.15 & -1.15~$\pm$~0.15  & S &B16 \\
J073149.50+285448 & 3.676 & 2.688 & 20.60~$\pm$~0.15 & -1.45~$\pm$~0.17 & -1.45~$\pm$~0.17  & Si & F14 \\
J0818+0958 & 3.656 & 3.306 & 21~$\pm$~0.1 & -1.51~$\pm$~0.1 & -1.51~$\pm$~0.1  & Si & B16 \\
J0920+0725 & 3.646 & 2.238 & 20.9~$\pm$~0.15 & -1.85~$\pm$~0.15 & -1.07~$\pm$~0.28  & Fe &B16 \\
J1020+0922 & 3.64 & 2.592 & 21.5~$\pm$~0.1 & -1.75~$\pm$~0.1 & -1.75~$\pm$~0.1  & Si & B16 \\
J034300.88-062229 & 3.623 & 2.571 & 20.75~$\pm$~0.10 & -2.29~$\pm$~0.26 & -1.99~$\pm$~0.35  & Fe &F14 \\
Q0201+11 & 3.61 & 3.387 & 21.26~$\pm$~0.10 & -1.20~$\pm$~0.15 & -1.20~$\pm$~0.15  & Si & R12 \\
J081618.99+482328 & 3.582 & 2.707 & 20.70~$\pm$~0.10 & -2.36~$\pm$~0.15 & -2.36~$\pm$~0.15  & Si & F14 \\
J115130.48+353625 & 3.581 & 2.598 & 20.90~$\pm$~0.10 & -1.28~$\pm$~0.11 & -1.28~$\pm$~0.11  & Si & F14 \\
J0013+1358 & 3.575 & 3.281 & 21.55~$\pm$~0.15 & -2.06~$\pm$~0.16 & -2.06~$\pm$~0.16  & Si & R12 \\
Q2223+20 & 3.561 & 3.119 & 20.30~$\pm$~0.10 & -2.18~$\pm$~0.11 & -2.18~$\pm$~0.11  & Si & R12 \\
J1517+0511 & 3.555 & 2.688 & 21.4~$\pm$~0.1 & -2.06~$\pm$~0.1 & -2.06~$\pm$~0.1  & Si & B16 \\
B1228-113 & 3.528 & 2.193 & 20.60~$\pm$~0.10 & -0.46~$\pm$~0.10 & 0.12~$\pm$~0.26  & Fe &R12 \\
J1024+1819 & 3.524 & 2.298 & 21.3~$\pm$~0.1 & -1.45~$\pm$~0.1 & -1.45~$\pm$~0.1  & Si & B16 \\
J0825+5127 & 3.51 & 3.318 & 20.85~$\pm$~0.10 & -1.67~$\pm$~0.14 & -1.67~$\pm$~0.14  & Si & R12 \\
J2238+0016 & 3.467 & 3.365 & 20.40~$\pm$~0.15 & -2.29~$\pm$~0.15 & -2.29~$\pm$~0.15  & Si & R12 \\
B0335-211 & 3.442 & 3.18 & 20.65~$\pm$~0.10 & -2.39~$\pm$~0.20 & -2.39~$\pm$~0.20  & Si & R12 \\
J095605.09+144854 & 3.435 & 2.661 & 20.85~$\pm$~0.10 & -1.46~$\pm$~0.12 & -1.46~$\pm$~0.12  & Si & F14 \\
J095604.43+344415 & 3.427 & 2.389 & 21.10~$\pm$~0.20 & -1.00~$\pm$~0.17 & -1.00~$\pm$~0.17  & Zn &F14 \\
Q0930+28 & 3.42 & 3.235 & 20.30~$\pm$~0.10 & -1.70~$\pm$~0.11 & -1.70~$\pm$~0.11  & Si & R12 \\
J0929+2825 & 3.404 & 3.263 & 21.10~$\pm$~0.10 & -1.57 & -1.57  & Si & R12 \\
J0929+2825 & 3.404 & 2.768 & 20.80~$\pm$~0.10 & -0.67~$\pm$~0.10 & -0.67~$\pm$~0.10  & Si & R12 \\
J2315+1456 & 3.39 & 3.273 & 20.30~$\pm$~0.15 & -1.68~$\pm$~0.16 & -1.68~$\pm$~0.16  & Si & R12 \\
J1200+4015 & 3.364 & 3.22 & 20.85~$\pm$~0.10 & -0.64~$\pm$~0.10 & -0.64~$\pm$~0.10  & Si & R12 \\
J132005.97+131015 & 3.352 & 2.672 & 20.30~$\pm$~0.20 & -2.30~$\pm$~0.10 & -2.30~$\pm$~0.10  & Si & F14 \\
J075155.10+451619 & 3.341 & 2.683 & 20.50~$\pm$~0.10 & -1.16~$\pm$~0.13 & -1.16~$\pm$~0.13  & Si & F14 \\
FJ2334-09 & 3.33 & 3.057 & 20.45~$\pm$~0.10 & -0.99~$\pm$~0.11 & -0.99~$\pm$~0.11  & Si & R12 \\
Q1209+0919 & 3.297 & 2.584 & 21.40~$\pm$~0.10 & -1.00~$\pm$~0.10 & -1.00~$\pm$~0.10  & Si & R12 \\
J0242-2917 & 3.27 & 2.56 & 20.90~$\pm$~0.10 & -1.94~$\pm$~0.11 & -1.94~$\pm$~0.11  & Si & R12 \\
Q1451+123 & 3.25 & 2.469 & 20.39~$\pm$~0.10 & -2.08~$\pm$~0.14 & -2.08~$\pm$~0.14  & Si & R12 \\
J2358+0149 & 3.235 & 2.979 & 21.69~$\pm$~0.10 & -1.71~$\pm$~0.26 & -1.71~$\pm$~0.26  & Si & S16 \\
J2358+0149 & 3.235 & 3.11 & 20.45~$\pm$~0.10 & -1.87~$\pm$~0.20 & -1.87~$\pm$~0.2  & O &S16 \\
J2358+0149 & 3.235 & 3.133 & 20.33 & -1.17~$\pm$~0.46 & -1.17~$\pm$~0.46  & Si & S16 \\
Q0347-38 & 3.23 & 3.025 & 20.63 & -1.11~$\pm$~0.03 & -1.11~$\pm$~0.03  & Si & R12 \\
HS0741+4741 & 3.22 & 3.017 & 20.48~$\pm$~0.10 & -1.62~$\pm$~0.10 & -1.62~$\pm$~0.10  & Si & R12 \\
eHAQ0111+0641 & 3.21 & 2.207 & 21.50~$\pm$~0.10 & -0.60~$\pm$~0.30 & -0.60~$\pm$~0.30  & Si & F17 \\
Q0135-273 & 3.21 & 2.8 & 21.00~$\pm$~0.10 & -1.52~$\pm$~0.13 & -1.52~$\pm$~0.13  & Si & R12 \\
J1410+5111 & 3.206 & 2.964 & 20.85~$\pm$~0.20 & -2.16~$\pm$~0.16 & -1.89~$\pm$~0.29  & Fe &R12 \\
J1410+5111 & 3.206 & 2.934 & 20.80~$\pm$~0.15 & -0.90~$\pm$~0.15 & -0.90~$\pm$~0.15  & Si & R12 \\
Q0336-01 & 3.2 & 3.062 & 21.20~$\pm$~0.10 & -1.49~$\pm$~0.10 & -1.49~$\pm$~0.10  & Si & R12 \\
Q1502+4837 & 3.2 & 2.57 & 20.30~$\pm$~0.15 & -1.58~$\pm$~0.17 & -1.58~$\pm$~0.17  & Si & R12 \\
Q1425+6039 & 3.17 & 2.827 & 20.30~$\pm$~0.04 & -0.99~$\pm$~0.05 & -0.50~$\pm$~0.24  & Fe &R12 \\
Q2138-444 & 3.17 & 2.852 & 20.80~$\pm$~0.08 & -1.72~$\pm$~0.13 & -1.45~$\pm$~0.27  & Fe &R12 \\
PKS1354-17 & 3.15 & 2.78 & 20.30~$\pm$~0.15 & -1.32~$\pm$~0.19 & -1.32~$\pm$~0.19  & Si & R12 \\
CTQ460 & 3.13 & 2.777 & 21.00~$\pm$~0.10 & -1.32~$\pm$~0.10 & -1.32~$\pm$~0.10  & Si & R12 \\
J1014+4300 & 3.125 & 2.959 & 20.50~$\pm$~0.10 & -0.50~$\pm$~0.10 & -0.50~$\pm$~0.10  & Si & R12 \\
BQ1021+3001 & 3.12 & 2.949 & 20.70~$\pm$~0.10 & -2.11~$\pm$~0.10 & -2.11~$\pm$~0.10  & Si & R12 \\
Q1008+36 & 3.119 & 2.799 & 20.70~$\pm$~0.05 & -1.62~$\pm$~0.06 & -1.62~$\pm$~0.06  & Si & R12 \\
J1240+1455 & 3.107 & 3.024 & 20.45~$\pm$~0.10 & -1.39~$\pm$~0.09 & -1.39~$\pm$~0.09  & Si & R12 \\
J0826+3148 & 3.09358 & 2.912 & 20.30~$\pm$~0.15 & -1.50~$\pm$~0.15 & -1.50~$\pm$~0.15  & Si & R12 \\
Q0058-292 & 3.09 & 2.671 & 21.10~$\pm$~0.10 & -1.39~$\pm$~0.13 & -1.39~$\pm$~0.13  & Si & R12 \\
J0642-5038 & 3.09 & 2.659 & 20.95~$\pm$~0.10 & -1.25~$\pm$~0.10 & -0.73~$\pm$~0.26  & Fe &R12 \\
B0913+003 & 3.074 & 2.743 & 20.74~$\pm$~0.10 & -1.06~$\pm$~0.17 & -0.57~$\pm$~0.29  & Fe &R12 \\
Q1759+75 & 3.05 & 2.625 & 20.76~$\pm$~0.01 & -0.67~$\pm$~0.01 & -0.67~$\pm$~0.01  & Si & R12 \\
Q0338-005 & 3.049 & 2.23 & 21.09~$\pm$~0.10 & -0.67~$\pm$~0.18 & -0.67~$\pm$~0.18  & Zn &B19 \\
Q0102-190 & 3.037 & 2.37 & 20.85~$\pm$~0.10 & -1.76~$\pm$~0.11 & -1.76~$\pm$~0.11  & Si & R12 \\
J0139-0824 & 3.02 & 2.678 & 20.70~$\pm$~0.15 & -1.15~$\pm$~0.19 & -1.15~$\pm$~0.19  & Si & R12 \\
CTQ247 & 3.02 & 2.622 & 20.47~$\pm$~0.10 & -1.97~$\pm$~0.10 & -1.97~$\pm$~0.10  & Si & R12 \\
CTQ247 & 3.02 & 2.55 & 21.13~$\pm$~0.10 & -1.46~$\pm$~0.12 & -1.46~$\pm$~0.12  & Si & R12 \\
CTQ247 & 3.02 & 2.595 & 21.09~$\pm$~0.10 & -1.05~$\pm$~0.11 & -1.05~$\pm$~0.11  & Si & R12 \\
Q2231-002 & 3.02 & 2.066 & 20.56~$\pm$~0.10 & -0.61~$\pm$~0.20 & -0.61~$\pm$~0.20  & Si & R12 \\
Q2348-01 & 3.014 & 2.615 & 21.30~$\pm$~0.10 & -1.93~$\pm$~0.12 & -1.93~$\pm$~0.12  & Si & R12 \\
Q2348-01 & 3.014 & 2.426 & 20.50~$\pm$~0.10 & -0.64~$\pm$~0.10 & -0.64~$\pm$~0.10  & Si & R12 \\
HE2243-6031 & 3.01 & 2.33 & 20.67~$\pm$~0.02 & -0.80~$\pm$~0.04 & -0.80~$\pm$~0.04  & Si & R12 \\
B1354-107 & 3.006 & 2.501 & 20.44~$\pm$~0.10 & -1.46~$\pm$~0.15 & -1.46~$\pm$~0.15  & Si & R12 \\
J040718.06-441014 &3 & 1.913 & 20.55~$\pm$~0.10 & -0.77~$\pm$~0.11 & -0.77~$\pm$~0.11  & Si & F14 \\
Q0216+08 & 2.992 & 2.293 & 20.50~$\pm$~0.10 & -0.61~$\pm$~0.10 & -0.61~$\pm$~0.10  & Si & R12 \\
J1035+5440 & 2.9884 & 2.684 & 20.50~$\pm$~0.20 & -0.86~$\pm$~0.16 & -0.28~$\pm$~0.28  & Fe &R12 \\
Q0112-306 & 2.985 & 2.418 & 20.50~$\pm$~0.08 & -2.40~$\pm$~0.10 & -2.40~$\pm$~0.10  & Si & R12 \\
Q0112-306 & 2.985 & 2.702 & 20.30~$\pm$~0.10 & -0.45~$\pm$~0.11 & -0.45~$\pm$~0.11  & Si & R12 \\
J1116+4118 & 2.982 & 2.663 & 20.48~$\pm$~0.10 & -0.94~$\pm$~0.16 & -0.94~$\pm$~0.16  & Si & R12 \\
Q2348-14 & 2.94 & 2.279 & 20.56~$\pm$~0.08 & -2.01~$\pm$~0.15 & -2.01~$\pm$~0.15  & Si & R12 \\
Q1223+17 & 2.92 & 2.466 & 21.50~$\pm$~0.10 & -1.54~$\pm$~0.10 & -1.54~$\pm$~0.10  & Si & R12 \\
Q1337+11 & 2.917 & 2.796 & 20.95~$\pm$~0.10 & -1.67~$\pm$~0.12 & -1.67~$\pm$~0.12  & Si & R12 \\
J1353+5328 & 2.9153 & 2.835 & 20.80~$\pm$~0.10 & -1.38~$\pm$~0.10 & -1.38~$\pm$~0.10  & Si & R12 \\
J2343+1410 & 2.913 & 2.677 & 20.50~$\pm$~0.15 & -2.21~$\pm$~0.17 & -1.94~$\pm$~0.29  & Fe &R12 \\
Q0201+36 & 2.91208 & 2.463 & 20.38~$\pm$~0.05 & -0.24~$\pm$~0.05 & -0.24~$\pm$~0.05  & Si & R12 \\
B0933-333 & 2.906 & 2.682 & 20.50~$\pm$~0.10 & -1.22~$\pm$~0.17 & -1.22~$\pm$~0.17  & Si & R12 \\
HS1132+2243 & 2.885 & 2.784 & 21.00~$\pm$~0.07 & -1.94~$\pm$~0.07 & -1.94~$\pm$~0.07  & Si & R12 \\
B0438-436 & 2.863 & 2.347 & 20.78~$\pm$~0.10 & -0.93~$\pm$~0.10 & -0.34~$\pm$~0.26  & Fe &R12 \\
Q1409+095 & 2.838 & 2.456 & 20.54~$\pm$~0.10 & -2.08~$\pm$~0.15 & -2.08~$\pm$~0.15  & Si & R12 \\
J1155+3510 & 2.8374 & 2.758 & 21.00~$\pm$~0.10 & -1.38~$\pm$~0.10 & -1.38~$\pm$~0.10  & Si & R12 \\
Q0112+029 & 2.81 & 2.423 & 20.78~$\pm$~0.08 & -1.24~$\pm$~0.11 & -1.24~$\pm$~0.11  & Si & R12 \\
Q2359-02 & 2.8 & 2.095 & 20.70~$\pm$~0.10 & -0.72~$\pm$~0.10 & -0.72~$\pm$~0.10  & Si & R12 \\
Q2359-02 & 2.8 & 2.154 & 20.30~$\pm$~0.10 & -1.53~$\pm$~0.10 & -1.53~$\pm$~0.10  & Si & R12 \\
Q2344+12 & 2.79 & 2.538 & 20.36~$\pm$~0.10 & -1.69~$\pm$~0.10 & -1.69~$\pm$~0.10  & Si & R12 \\
Q0528-2505 & 2.78 & 2.141 & 20.95~$\pm$~0.05 & -1.27~$\pm$~0.06 & -1.27~$\pm$~0.06  & Si & R12 \\
J1241+4617 & 2.7697 & 2.667 & 20.70~$\pm$~0.10 & -1.84~$\pm$~0.15 & -1.84~$\pm$~0.15  & Si & R12 \\
Q2343+125 & 2.763 & 2.431 & 20.34~$\pm$~0.10 & -0.58~$\pm$~0.10 & -0.58~$\pm$~0.10  & Si & R12 \\
J0812+3208 & 2.701 & 2.626 & 21.35~$\pm$~0.10 & -0.87~$\pm$~0.13 & -0.87~$\pm$~0.13  & Si & R12 \\
Q0836+11 & 2.7 & 2.465 & 20.58~$\pm$~0.10 & -1.10~$\pm$~0.11 & -1.10~$\pm$~0.11  & Si & R12 \\
PH957 & 2.69 & 2.309 & 21.37~$\pm$~0.08 & -1.41~$\pm$~0.08 & -1.41~$\pm$~0.08  & Si & R12 \\
Q0449-1645 & 2.679 & 1.007 & 20.98~$\pm$~0.07 & -0.96~$\pm$~0.08 & -0.96~$\pm$~0.08  & Zn &L13 \\
B0432-440 & 2.65 & 2.302 & 20.95~$\pm$~0.10 & -1.18~$\pm$~0.15 & -1.81~$\pm$~0.15  & Si & R12 \\
Q1215+33 & 2.61 & 1.999 & 20.95~$\pm$~0.07 & -1.43~$\pm$~0.07 & -1.43~$\pm$~0.07  & Si & R12 \\
J2036-0553 & 2.58 & 2.28 & 21.20~$\pm$~0.15 & -1.81~$\pm$~0.15 & -1.81~$\pm$~0.15  & Si & R12 \\
J1238+3437 & 2.5718 & 2.471 & 20.80~$\pm$~0.10 & -2.04~$\pm$~0.15 & -2.04~$\pm$~0.15  & Si & R12 \\
Q1232+08 & 2.57 & 2.337 & 20.90~$\pm$~0.10 & -1.22~$\pm$~0.15 & -1.22~$\pm$~0.15  & Si & R12 \\
J1208+6303 & 2.561 & 2.444 & 20.70~$\pm$~0.15 & -2.28~$\pm$~0.15 & -2.28~$\pm$~0.15  & Si & R12 \\
Q2206-19 & 2.56 & 1.92 & 20.65~$\pm$~0.07 & -0.38~$\pm$~0.07 & -0.38~$\pm$~0.07  & Si & R12 \\
Q2206-19 & 2.56 & 2.076 & 20.43~$\pm$~0.06 & -2.26~$\pm$~0.07 & -2.26~$\pm$~0.07  & Si & R12 \\
J1541+3153 & 2.553 & 2.443 & 20.95~$\pm$~0.10 & -1.49~$\pm$~0.11 & -1.49~$\pm$~0.11  & Si & R12 \\
Q1210+17 & 2.54 & 1.892 & 20.60~$\pm$~0.10 & -0.79~$\pm$~0.10 & -0.79~$\pm$~0.10  & Si & R12 \\
B1055-301 & 2.523 & 1.903 & 21.54~$\pm$~0.10 & -1.50~$\pm$~0.10 & -1.09~$\pm$~0.26  & Fe &R12 \\
KV-RQ1500-0031 & 2.52 & 2.428 & 21.20~$\pm$~0.14 & -0.90~$\pm$~0.20 & -0.90~$\pm$~0.20  & Si & H18 \\
J2151-0707 & 2.52 & 2.327 & 20.45~$\pm$~0.15 & -1.60~$\pm$~0.15 & -1.60~$\pm$~0.15  & Si & R12 \\
J2141+1119 & 2.51 & 2.426 & 20.30~$\pm$~0.20 & -1.92~$\pm$~0.20 & -1.92~$\pm$~0.20  & Si & R12 \\
Q0841+12 & 2.5 & 2.374 & 20.95~$\pm$~0.09 & -1.33~$\pm$~0.09 & -1.33~$\pm$~0.0  & Si & R12 \\
J084424.24+124546 & 2.482 & 1.864 & 21.00~$\pm$~0.10 & -1.54~$\pm$~0.12 & -1.54~$\pm$~0.12  & Si & F14 \\
J0039-3354 & 2.48 & 2.224 & 20.60~$\pm$~0.10 & -1.26~$\pm$~0.12 & -1.26~$\pm$~0.12  & Si & R12 \\
B2314-409 & 2.448 & 1.857 & 20.90~$\pm$~0.10 & -0.95~$\pm$~0.20 & -0.95~$\pm$~0.20  & Si & R12 \\
Q0149+33 & 2.43 & 2.141 & 20.50~$\pm$~0.10 & -1.44~$\pm$~0.11 & -1.44~$\pm$~0.11  & Si & R12 \\
Q1122-168 & 2.4 & 0.682 & 20.45~$\pm$~0.05 & -1.00~$\pm$~0.15 & -1.00~$\pm$~0.15  & Zn &L13,R12 \\
B1230-101 & 2.394 & 1.931 & 20.48~$\pm$~0.10 & -0.42~$\pm$~0.11 & 0.16~$\pm$~0.26  & Fe &R12 \\
J0300-3152 & 2.37 & 2.179 & 20.80~$\pm$~0.08 & -1.75~$\pm$~0.09 & -1.75~$\pm$~0.09  & Si & R12 \\
Q0551-366 & 2.318 & 1.962 & 20.50~$\pm$~0.08 & -0.27~$\pm$~0.15 & -0.27~$\pm$~0.1  & Si & R12 \\
Q1104-18 & 2.31 & 1.661 & 20.80~$\pm$~0.10 & -0.99~$\pm$~0.10 & -0.99~$\pm$~0.10  & Si & R12 \\
Q0450-13 & 2.3 & 2.067 & 20.53~$\pm$~0.08 & -1.39~$\pm$~0.09 & -1.39~$\pm$~0.09  & Si & R12 \\
Q0235+164 & 2.29 & 0.524 & 21.65~$\pm$~0.15 & -0.58~$\pm$~0.15 & -0.58~$\pm$~0.15  & Zn &L13,R12 \\
Q0458-02 & 2.29 & 2.04 & 21.65~$\pm$~0.09 & -1.39~$\pm$~0.09 & -0.87~$\pm$~0.26  & Fe &R12 \\
J0254-4025 & 2.28 & 2.046 & 20.45~$\pm$~0.10 & -1.50~$\pm$~0.11 & -1.50~$\pm$~0.11  & Si & R12 \\
J0421-2624 & 2.28 & 2.157 & 20.65~$\pm$~0.10 & -1.81~$\pm$~0.10 & -1.81~$\pm$~0.10  & Si & R12 \\
J0049-2820 & 2.256 & 2.071 & 20.45~$\pm$~0.10 & -1.26~$\pm$~0.12 & -1.26~$\pm$~0.12  & Si & R12 \\
J2228-3954 & 2.21 & 2.095 & 21.20~$\pm$~0.10 & -1.56~$\pm$~0.11 & -1.15~$\pm$~0.26  & Fe &R12 \\
Q2230+02 & 2.15 & 1.864 & 20.85~$\pm$~0.08 & -0.71~$\pm$~0.09 & -0.71~$\pm$~0.09  & Si & R12 \\
Q0010-002 & 2.145 & 2.025 & 20.80~$\pm$~0.10 & -1.40~$\pm$~0.12 & -0.99~$\pm$~0.27  & Fe &R12 \\
TXS1755+578 & 2.11 & 1.97 & 21.4 & -0.25~$\pm$~0.19 & -0.25~$\pm$~0.19  & Zn &K14 \\
TXS1850+402 & 2.11 & 1.989 & 21.30~$\pm$~0.10 & -0.68~$\pm$~0.04 & -0.68~$\pm$~0.04  & Zn &K14 \\
Q0013-004 & 2.086 & 1.973 & 20.83~$\pm$~0.07 & -0.70~$\pm$~0.08 & -0.70~$\pm$~0.08  & Si & R12 \\
Q1331+17 & 2.08 & 1.776 & 21.14~$\pm$~0.08 & -1.37~$\pm$~0.08 & -1.37~$\pm$~0.08  & Si & R12 \\
Q1354+258 & 2.006 & 1.42 & 21.54~$\pm$~0.06 & -1.69~$\pm$~0.13 & -1.69~$\pm$~0.13  & Si & R12 \\
Q0935+417 & 1.98 & 1.373 & 20.52~$\pm$~0.10 & -1.14~$\pm$~0.13 & -0.61~$\pm$~0.25  & Fe &R12 \\
Q1501+0019 & 1.93 & 1.483 & 20.85~$\pm$~0.13 & -0.65~$\pm$~0.13 & -0.65~$\pm$~0.13  & Si & R12 \\
Q0933+732 & 1.89 & 1.479 & 21.62~$\pm$~0.10 & -1.77~$\pm$~0.10 & -1.50~$\pm$~0.26  & Fe &R12 \\
Q0948+433 & 1.89 & 1.233 & 21.62~$\pm$~0.05 & -1.33~$\pm$~0.05 & -0.81~$\pm$~0.25  & Fe &R12 \\
Q1007+0042 & 1.681 & 1.037 & 21.15~$\pm$~0.24 & -0.51~$\pm$~0.24 & -0.51~$\pm$~0.24  & Zn &L13,R12 \\
Q2149+212 & 1.538 & 0.911 & 20.70~$\pm$~0.10 & -0.93 & -0.93  & Zn &L13 \\
Q1224+0037 & 1.482 & 1.235 & 20.88~$\pm$~0.05 & -1.29~$\pm$~0.09 & -1.29~$\pm$~0.09  & Si & R12 \\
Q0952+179 & 1.472 & 0.238 & 21.32~$\pm$~0.05 & -1.78~$\pm$~0.24 & -1.51~$\pm$~0.34  & Fe &R12 \\
Q1727+5302 & 1.444 & 0.945 & 21.16~$\pm$~0.05 & -0.56~$\pm$~0.06 & -0.56~$\pm$~0.06  & Zn &L13,R12 \\
Q1727+5302 & 1.444 & 1.031 & 21.41~$\pm$~0.03 & -1.36~$\pm$~0.08 & -1.36~$\pm$~0.08  & Zn &L13,R12 \\
Q0302-223 & 1.4 & 1.009 & 20.36~$\pm$~0.11 & -0.73~$\pm$~0.12 & -0.73~$\pm$~0.12  & Zn &L13 \\
Q1010+0003 & 1.398 & 1.265 & 21.52~$\pm$~0.07 & -1.43~$\pm$~0.09 & -1.02~$\pm$~0.26  & Fe &R12 \\
J1107+0048 & 1.392 & 0.74 & 21.00~$\pm$~0.03 & -0.54~$\pm$~0.20 & -0.54~$\pm$~0.20  & Zn &L13,R12 \\
Q1323-0021 & 1.39 & 0.716 & 20.54~$\pm$~0.15 & 0.12~$\pm$~0.26 & 0.12~$\pm$~0.26  & Zn &L13 \\
Q0139-0023 & 1.384 & 0.682 & 20.60~$\pm$~0.12 & 0.08 & 0.08  & Zn &L13 \\
Q0454+039 & 1.35 & 0.86 & 20.69~$\pm$~0.06 & -0.79~$\pm$~0.12 & -0.79~$\pm$~0.12  & Zn &L13,R12 \\
J0256+0110 & 1.348 & 0.725 & 20.70~$\pm$~0.15 & -0.11~$\pm$~0.04 & -0.11~$\pm$~0.04  & Zn &L13,R12 \\
J2328+0022 & 1.309 & 0.652 & 20.32~$\pm$~0.07 & -0.49~$\pm$~0.22 & -0.49~$\pm$~0.22  & Zn &L13,R12 \\
Q1225+0035 & 1.226 & 0.773 & 21.38~$\pm$~0.12 & -0.78~$\pm$~0.14 & -0.78~$\pm$~0.14  & Zn &L13,R12 \\
J1431+3952 & 1.215 & 0.602 & 21.20~$\pm$~0.10 & -0.80~$\pm$~0.20 & -0.80~$\pm$~0.20  & Zn &L13 \\
Q1733+5533 & 1.074 & 0.998 & 20.70~$\pm$~0.04 & -0.44~$\pm$~0.07 & -0.44~$\pm$~0.07  & Zn &L13,R12 \\
Q1229-021 & 1.043 & 0.395 & 20.75~$\pm$~0.07 & -0.47~$\pm$~0.15 & -0.47~$\pm$~0.15  & Zn &L13 \\
Q0253+0107 & 1.035 & 0.632 & 20.78~$\pm$~0.12 & -0.35~$\pm$~0.35 & -0.35~$\pm$~0.35  & Zn &L13 \\
Q1137+3907 & 1.023 & 0.72 & 21.10~$\pm$~0.10 & -0.54~$\pm$~0.11 & 0.05~$\pm$~0.26  & Fe &R12 \\
Q0827+243 & 0.941 & 0.525 & 20.30~$\pm$~0.04 & -1.10~$\pm$~0.16 & -0.61~$\pm$~0.29  & Fe &R12 \\
Q1622+238 & 0.927 & 0.656 & 20.36~$\pm$~0.10 & -0.87~$\pm$~0.25 & -0.87~$\pm$~0.25  & Zn &L13,R12 \\
3C196 & 0.87 & 0.437 & 20.36~$\pm$~0.05 & -1.58~$\pm$~0.10 & -1.58~$\pm$~0.10  & Si & R12 \\
Q1328+307 & 0.849 & 0.692 & 21.25~$\pm$~0.06 & -1.20~$\pm$~0.09 & -1.20~$\pm$~0.09  & Zn &L13,R12 \\
J2353-0028 & 0.765 & 0.604 & 21.54~$\pm$~0.15 & -0.92~$\pm$~0.32 & -0.92~$\pm$~0.32  & Zn &L13,R12 \\
J0951+3307 & 0.6499 & 0.005 & 21 & -0.99~$\pm$~0.14 & -0.99~$\pm$~0.14  & S &K18 \\
Q0738+313 & 0.635 & 0.221 & 20.90~$\pm$~0.07 & -0.94~$\pm$~0.18 & -0.36~$\pm$~0.30  & Fe &R12 \\
Q0738+313 & 0.635 & 0.091 & 21.18~$\pm$~0.05 & -1.56~$\pm$~0.24 & -1.14~$\pm$~0.34  & Fe &R12 \\
J1619+3342 & 0.471 & 0.096 & 20.55~$\pm$~0.10 & -0.63~$\pm$~0.13 & -0.63~$\pm$~0.13  & S &L13 \\
J1009+0713 & 0.456 & 0.114 & 20.68~$\pm$~0.10 & -0.58~$\pm$~0.16 & -0.58~$\pm$~0.16  & S &L13,R12 \\
J1616+4154 & 0.44 & 0.321 & 20.60~$\pm$~0.20 & -0.35~$\pm$~0.23 & -0.35~$\pm$~0.23  & S &L13 \\
B0120-28 & 0.434 & 0.186 & 20.5 & -1.19~$\pm$~0.21 & -1.19~$\pm$~0.21  & S &O14 \\
J1512+0128 & 0.2654 & 0.029 & 20.4 & -1.29~$\pm$~0.13 & -1.29~$\pm$~0.13  & Si & K18\\
\enddata
\end{deluxetable*}

\appendix
\setcounter{figure}{0} 
\renewcommand{\thefigure}{A\arabic{figure}}
\setcounter{table}{0}
\renewcommand{\thetable}{A\arabic{table}}

\section{Dust Depletion and $\alpha$-Enhancement Correction Using Zinc}

\label{sec: appendix_a}
An alternate option for the dust depletion method that we describe and use in Sec. \ref{sec: dust corrections} is to use Zn instead of S to obtain the model for correction. Zn is similarly observed to behave like S (see R12). However, since there is less agreement in the literature about the true nature of Zn (see also R12), employing this element for a dust correction is less straightforward. Here, we show that the use of Zn for the depletion correction does not affect the results of this paper. 

\begin{figure*}[b!]
    \centering
    \includegraphics[width = 7.1 in]{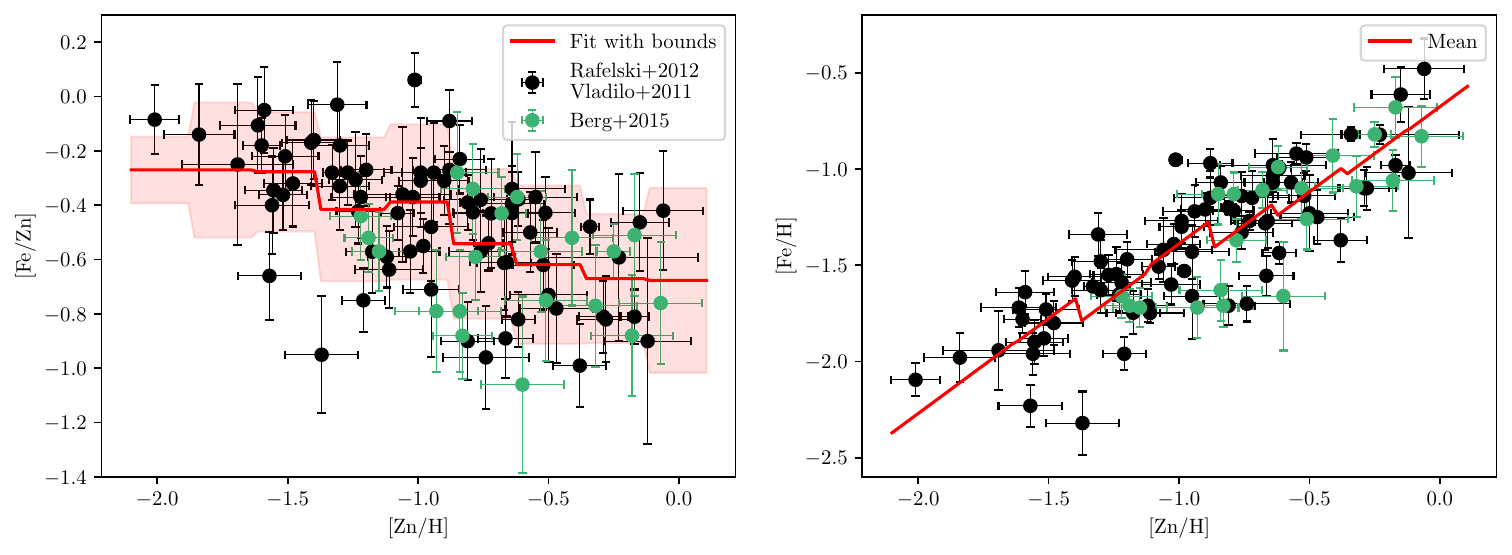}
    \caption{{\emph Left:} Measured relationship between [Zn/H] and [Fe/Zn] using the same studies as in Fig. \ref{fig:s_fe}. The vertical and horizontal error bars are the uncertainties on the respective measures. After binning the data, the mean relationship between [Zn/H] and [Fe/Zn] is shown in red, with uncertainties shown as red shading. {\emph Right:} Parameterized relationship from the left panel fit to [Zn/H] vs [Fe/H] parameter space.}
    \label{fig:zn_fe}
\end{figure*}
Similarly to our method to correct using [S/H], we employ a data binning approach that characterizes the relationship between [Fe/Zn] and [Zn/H], seen in the left panel of Fig. \ref{fig:zn_fe}. We also apply this relationship in the gas-phase abundance space, as seen in the right panel of the figure. We apply this model between to all DLAs in our sample with metallicities obtained via Fe transition lines. This yields  
\begin{equation}
\langle Z \rangle = (-0.22 \pm 0.05)\times z - (0.50 \pm 0.14),
\end{equation}
with a $\langle Z \rangle$ value of $\langle Z \rangle = -2.00^{+0.10}_{-0.12}$ for the highest redshift bin. This column density weighted metallicity value for the high-z bin is consistent with what we calculate using the sulfur-based corrections. We measure a $4.4\sigma$ deviation of the highest redshift bin from the value expected by the fit to the rest of the sample and a p-value of 0.03 from the K-S test. This indicates that we can reject the hypothesis that the the low redshift sample and the high redshift sample are from the same distribution at $3\sigma$ confidence. This is consistent with the confidence level yielded in our main results using the S based dust depletion and $\alpha$-enhancement correction. 

\setcounter{figure}{0} 
\renewcommand{\thefigure}{B\arabic{figure}}
\setcounter{table}{0}
\renewcommand{\thetable}{B\arabic{table}}
\setcounter{equation}{0}
\renewcommand{\theequation}{B\arabic{equation}}

\section{Variations on the Cosmic Metallicity Evolution}
\label{sec: appendix_b}

\begin{figure}
    \centering
    \includegraphics[scale = 0.6]{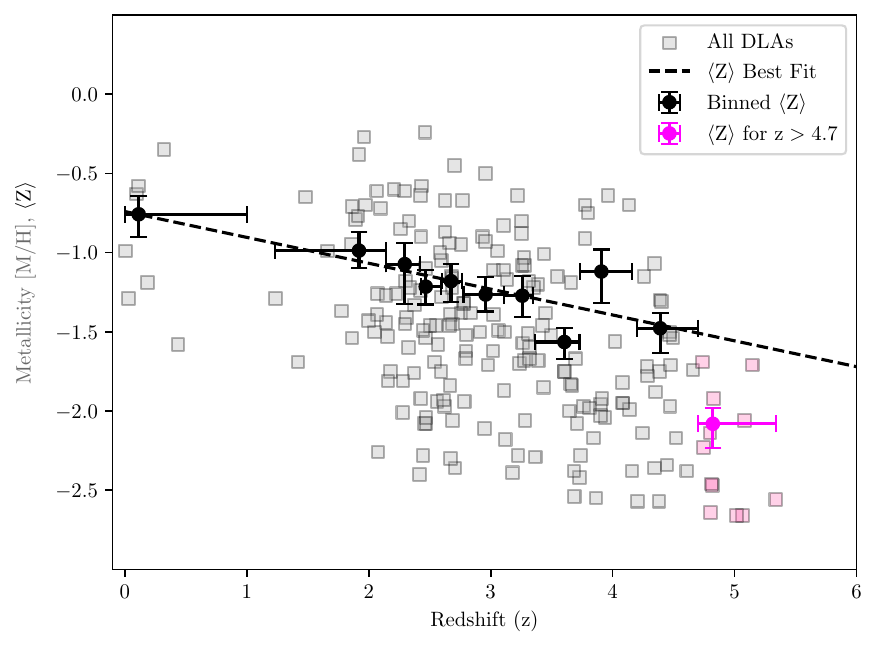}
    \includegraphics[scale = 0.6]{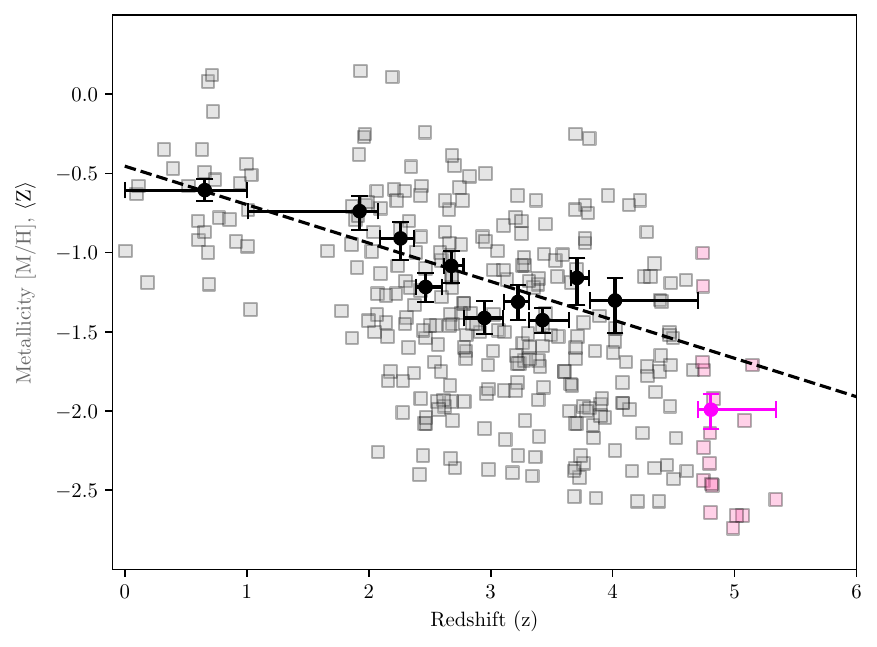}
    \caption{\emph{Left:} Same as right panel of Fig. \ref{fig:metal z trend}, but with all DLAs with a reported metallicity using Fe or Zn abundances removed from the sample. The linear trend shown by the black dashed line is now fit to the reduced sample. \emph{Right:} Same as left panel here but with all DLAs with reported metallicities using Fe or Si abundances at $z < 1.5$ removed from the sample instead.}
    \label{fig: noFeZn}
\end{figure}

Given the uncertainties of $\alpha$-enhancement and dust depletion correction, we consider the evolution with only $\alpha$-elements. The $\alpha$-elements we use in this work have relatively low depletion, so this provides a good test of our methodology of utilizing a mix of element types (e.g. Fe or Zn based abundances), which is preferred due to larger number statistics. As we apply a combined dust depletion and $\alpha$-enhancement correction, we remove DLAs across all redshifts that use either Fe or Zn abundances to compute the reported metallicity that we use in this work, reducing our overall sample from 306 DLAs to 202 DLAs. We force the lowest bin to span only DLAs $z< 1$ to avoid fitting roughly two-thirds the age of the universe in one bin. We keep the intermediate redshift bins to have an equal number of observations per bin. We keep all other methods to our analysis and computation of the cosmic evolution of metallicity identical, and we show the results of this in left panel of Fig. \ref{fig: noFeZn}. 
With these modifications to the sample, the linear trend fit to the $z < 4.7$ becomes:
\begin{equation}
    \langle Z \rangle = (-0.16 \pm 0.04)\times z - (0.74 \pm 0.11),
\end{equation}
which is consistent with the main trend reported in Section \ref{sec: metal evolution}. Additionally, the deviation of the binned data point for DLAs $z > 4.7$ is $5.0\sigma$ deviant from the value of that bin as predicted by the $z < 4.7$ linear trend. These findings are consistent with the results presented in the main text. Thus, we conclude that the turnover in the metal content of the universe at $z > 4.7$ is not caused by the inclusion of Fe- or Zn-based metallicities in the analysis and the measured slope is robust. 

We additionally test the impact of our dust depletion and $\alpha$-enhancement correction on the low-z DLAs, as this population tends to have higher metallicity values that are most impacted by the depletion of dust. We remove the DLAs at $z< 1.5$ where we use the Fe or Si abundances; however we keep the S and Zn measurements as we need a population of $z< 1.5$ DLAs to act as a lever arm for the lower end of our linear fit and these are less affected by dust depletion. We also fix the lowest redshift bin here to span $z< 1$, while keeping the intermediate redshift bins to have an equal number of observations per bin. We show the results of the fit in the right panel of Fig. \ref{fig: noFeZn}. The linear trend resulting from this modified sample of DLAs is: 
\begin{equation}
    \langle Z \rangle = (-0.24 \pm 0.05)\times z - (0.45 \pm 0.16),
\end{equation}
and the deviation of the $z > 4.7$ group from the predicted value based on the low-z population remains $4.1\sigma$, consistent with the results presented in Section \ref{sec: metal evolution}. Therefore, the turnover in metallicity at high-z and slope in this work is independent of these higher low-z depletion levels as the linear fit is not significantly impacted by their removal.

\setcounter{figure}{0} 
\renewcommand{\thefigure}{C\arabic{figure}}
\setcounter{table}{0}
\renewcommand{\thetable}{C\arabic{table}}
\section{Literature Search Sources}
\label{sec: appendix_c}

We consider many more publications from the literature than those that made it into our final unbiased in metallicity based sample presented in the main body of this work. Table \ref{tab: all papers} lists all publications considered over the course of this work, with a cutoff deadline of August 15, 2025 and a flag explaining why some or all DLAs from each publication were not included in our final sample. 

\begin{table*}
\centering
\begin{tabular}{lc||lc}
Publication & Flag & Publication & Flag\\
\hline
\citet{banados2019} & 2 & \citet{kulkarni2015} & 1 \\
\citet{Bashir2019} & 0 & \citet{Lehner2013} & 0 \\
\citet{Berg2016} & *  & \citet{lin2023} & 4 \\
\citet{boettcher2021} & 2 & \citet{mackenzie2019} & 5 \\
\citet{boettcher2022} & 2 & \citet{mawatari2016} & 4 \\
\citet{Borthakur2019} & 4 & \citet{moller2018} & 1 \\
\citet{christenson2019} & 3 & \citet{morrison2016} & 3 \\
\citet{cooke2016} & 1 & \citet{neeleman2016} & 4 \\
\citet{cooke2017} & 1 & \citet{neilsen2022} & 2 \\
\citet{cooper2019} & 3 & \citet{noterdaeme2012} & 1 \\
\citet{DOdorico2018} & 2 & \citet{Oliveira2014} & 0 \\
\citet{DeCia2016} & 1 & \citet{pan2017} & * \\
\citet{decia2018} & 5 & \citet{Poudel2018} & * \\
\citet{dupuis2021} & 4  & \citet{Poudel2020} & * \\
\citet{fathivavsari2017} & 2 & \citet{poudel2021} & 3 \\
\citet{Fumagalli2014} & * & \citet{Rafelski2012} & 0 \\
\citet{fynbo2023} & 4 & \citet{Rafelski2014} & 0 \\
\citet{Heintz2018} & 0 & \citet{srianand2014} & 5 \\
\citet{Huyan2025} & * & \citet{Srianand2016} & *\\
\citet{Kanekar2014} & 0 & \citet{wang2015} & 5  \\
\citet{Kanekar2018} & * & \citet{welsh2019} & 1\\
\citet{kaur2022} & 1 & \citet{welsh2022} & 1\\
\citet{krogager2016} & 0 & \citet{welsh2023} & 1 \\
\end{tabular}

\caption{Full list of publications considered for inclusion in our sample. The names of publications that were included are given a flag of 0. The flags in columns two and four reference if DLAs from the publication were included in our sample or the reason why DLAs from the publication were excluded from our sample (1 = biased, 2 = proximate, 3 = sub-DLA or non-DLA, 4 = no/poor metallicity calculation, 5 = duplicates, and * = a combination of numerical flags). For detailed notes about starred citations, see Section \ref{sec: lit search}, as we include and exclude a combination of DLAs from those publications.}
\label{tab: all papers}
\end{table*}

\begin{acknowledgements}
This material is based upon work supported by the National Science Foundation under grant No. 2107989. This project was supported in part by the NSF grant AST-2107990 and by the Nantucket Maria Mitchell Association. MR also acknowledges support from the STScI's Director's Discretionary Research Funding (grant ID D0101.90228). The W. M. Keck Observatory is operated as a scientific partnership among the California Institute of Technology, the University of California, and the National Aeronautics and Space Administration. The Observatory was made possible by the generous financial support of the W. M. Keck Foundation. The authors wish to recognize and acknowledge the very significant cultural role and reverence that the summit of Mauna Kea has always had with the indigenous Hawaiian community.
\end{acknowledgements}

\newpage
\bibliographystyle{aasjournal}
\bibliography{refs.bib}
\vspace{5mm}
\facilities{Keck{ESI, HIRES}, VLT{X-shooter}}
\software{astropy \citep{2013A&A...558A..33A,2018AJ....156..123A}, Linetools \citep{linetools}, pyigm \citep{pyigm}
  }

\end{document}